\documentclass[%
aip,
rsi,%
amsmath,amssymb,
reprint,%
floatfix,
]{revtex4-1}

\usepackage{graphicx}
\usepackage{dcolumn}
\usepackage{braket}
\usepackage{bm}
\usepackage{multirow}
\usepackage{csquotes}
\usepackage{amsmath}
\usepackage{amssymb}
\usepackage{multirow}

\usepackage{xcolor}

\renewcommand{\eqref}[1]{eq.~(\ref{#1})}
\newcommand{\kt}{k_\mathrm{B}T}
\newcommand{\rmd}{\ensuremath{\mathrm{d}}}
\newcommand{\bx}{\ensuremath{\mathbf{x}}}
\newcommand{\bp}{\ensuremath{\mathbf{p}}}
\newcommand{\bq}{\ensuremath{\mathbf{q}}}
\newcommand{\kb}{\ensuremath{k_\mathrm{B}}}
\newcommand{\mwc}{\ensuremath{{\widetilde{\mathbf{x}}}}}

\begin{document}
	

\title{Entropy and Energy Profiles of Chemical Reactions}

\author{Johannes C. B. Dietschreit}
\affiliation{Department of Materials Science and Engineering, Massachusetts Institute of Technology, Cambridge, Massachusetts 02139, USA}

\author{Dennis J. Diestler}
\affiliation
{University of Nebraska-Lincoln, Lincoln, Nebraska 68583, USA}

\author{Rafael G\'{o}mez-Bombarelli}
\email{rafagb@mit.edu}
\affiliation{Department of Materials Science and Engineering, Massachusetts Institute of Technology, Cambridge, Massachusetts 02139, USA}

\date{20.4.2023}

		\begin{abstract}
			The description of chemical processes at the molecular level is often facilitated by use of reaction coordinates, or collective variables (CVs). 
			The CV measures the progress of the reaction and allows the construction of profiles that track the evolution of a specific property as the reaction progresses.
			Whereas CVs are routinely used, especially alongside enhanced sampling techniques, links between profiles and thermodynamic state functions and reaction rate constants are not rigorously exploited.   
			Here, we report a unified treatment of such reaction profiles. 
			Tractable expressions are derived for the free-energy, internal-energy, and entropy profiles as functions of only the CV.
			We demonstrate the ability of this treatment to extract quantitative insight from the entropy and internal-energy profiles of various real-world physicochemical processes, including intramolecular organic reactions, ionic transport in superionic electrolytes, and  molecular transport in nanoporous materials. 
			
		\end{abstract}

		\maketitle
		
		\section{Introduction}
		Computational analysis of chemical reactions at the atomic level allows mechanistic insights often inaccessible to experiments, as well as predictions of interesting or useful phenomena. 
		The statistical-mechanical treatment of reactions, which dates back decades, has produced tremendous scientific knowledge.\cite{Zwanzig1954, Kollman1993, Pande1997, Roux2004, Ludovic2011, elber_2017, Dietschreit2019, Naydenova2019, VonDerEsch2019, Roux2021, Dietschreit2022vdE, Henin2022, Axelrod2022} 
		Nevertheless, kinetics, thermodynamic state functions, and the evolution of properties over the course of the reaction 
		have not been directly and unambiguously connected in the literature. 
		Our purpose here is to define and exploit these relationships.
		
		From a microscopic viewpoint, a chemical reaction involves the atomic rearrangement from one depression on the potential energy surface (PES) to another.
		These minima are separated by a barrier that is overcome as the transition takes place. 
		The course of the reaction can be described by a \enquote{collective variable} (CV) (or \enquote{reaction coordinate}) usually defined as a function of the Cartesian coordinates of the system.\cite{Peters2016, Rogal2021} 
		The CV, which  provides a measure of progress of the reaction, must be chosen judiciously so that it has two non-overlapping domains corresponding to the regions of configuration space that define reactant and product. 
		A plethora of methods designed to determine the CV have been developed.\cite{Wu2017b, McGibbon2017, Sultan2017, Chen2018, Mendels2018a, Ribeiro2018, Wang2019d, Sun2020, Bonati2020, Wang2021, Hooft2021, Novelli2022, Sipka2023}
		However, for a complex realistic system it is almost impossible to find an optimal CV.\cite{Bolhuis2000} 
		
		The systematic characterization of the reaction is facilitated by introduction of the concept of \enquote{profile}, which is a plot of a particular property of the system obtained as an ensemble average over the system with the CV fixed at the value $z$. 
		The profile, which is  a function of only $z$, shows how the property varies as the reaction progresses. 
		The most sought after profile is the potential of mean force (PMF).\cite{Valleau1977, Darve2001, Laio2002, Darve2008, Fiorin2013, Comer2015, Fu2016, lesage2017smoothed} 
		The PMF is proportional to the negative  logarithm of the probability density on the CV and often called the free-energy profile. 
		However, as this profile is not  independent of the gauge\cite{hartmann2011two} (i.e., differences along the PMF depend on the specific formulation of the CV), it cannot be the free energy of the system.
		A rigorous, gauge-invariant expression for the free energy of a constrained system is thus lacking, as are expressions for other properties, such as entropy or internal energy of reaction and activation.
		
		We derive such expressions here both for the reaction overall and for the activation step, based on our previous work.\cite{DietschreitDiestlerOchsenfeld2022, DietschreitDiestlerBombarelli2022}
		These expressions are then applied to several diverse, realistic chemical systems: 
		i) an intramolecular radical reaction, ii) ion hopping in a solid electrolyte that exhibits correlated superionic conduction, and iii) migration of the [Cu(NH$_3$)$_2$]$^+$ catalyst complex in chabazite, a nanoporous aluminosilicate used in catalytic converters. 
		We are then able to trace the entropy and internal-energy profiles and observe features, such as local traps and barriers on both profiles than can be related to the behavior of each particular system.

		\section{Theory}

		\subsection{Preliminaries}
		
		Our proposed approach builds on the framework introduced in  Refs.~\citenum{DietschreitDiestlerOchsenfeld2022, DietschreitDiestlerBombarelli2022}, which is summarized in this subsection.  
		We treat the system as a molecular entity consisting of $N$ atoms. 
		The PES has two distinct minima corresponding to reactant (R) and product (P), and we regard the transition between those as the chemical reaction
		\begin{equation}
			\mathrm{R} \rightleftharpoons \mathrm{P}\ .
			\label{eq:reaction_RP}
		\end{equation}
		The species $\alpha$ (= R , P) is defined by the portion of 3$N$-dimensional configuration space it occupies, which is signified by $\Omega_\alpha$. 
		We assume that $\Omega_\mathrm{R}$ and $\Omega_\mathrm{P}$ constitute the whole configuration space. 
		We further assume that they are separated by a ($3N-1$)-dimensional dividing surface, which is taken to be the ridge of the potential barrier of the PES that separates the minima. 
		The configuration of the atoms is specified by $\bx = (x_1,\ x_2,\ \dots,\ x_{3N})^\mathrm{T}$, the column vector of Cartesian coordinates of the $N$ atoms; $\bp = (p_{1},\ p_{2},\ \dots,\ p_{3N})^\mathrm{T}$ is the column vector of corresponding conjugate momenta. 
		The Hamiltonian is expressed as
		\begin{equation}
			\mathcal{H} 
			= \frac{1}{2} \bp^\mathrm{T} \mathbf{M}^{-1} \bp +  U(\bx)
			= K(\bp) + U(\bx) \ ,
			\label{eq:Hamiltonian_full}
		\end{equation}
		where $\mathbf{M}$ stands for the diagonal matrix of atomic masses ($[\mathbf{M}]_{ii} = m_i$), $U(\bx)$ for the PES, and $K(\bp)$ for the kinetic energy.
		
		The canonical molecular partition function of species $\alpha$ is given by
		\begin{equation}
			q_\alpha = \frac{1}{h^{3N}} \int_{\Omega_\alpha} \rmd \bx\ \int \rmd \bp\ e^{-\beta \mathcal{H}} \ ,
			\label{eq:Q_molec}
		\end{equation}
		where $\rmd \bx = \prod_{i=1}^{3N} \rmd x_i$, $\int \rmd \bp = \prod_{i=1}^{3N} \int_{-\infty}^{\infty} \rmd p_i$, $\beta \equiv 1/ \kt$, $k_\mathrm{B}$ is Boltzmann's constant, $T$ is the absolute temperature, and $h$ is Planck's constant.
		Using \eqref{eq:Hamiltonian_full}, we perform the integrations on momenta in \eqref{eq:Q_molec} to get
		\begin{equation}
			q_\alpha = \frac{Z_\alpha}{\Lambda} \ ,
			\label{eq:Q_molec_integrated}
		\end{equation}
		where
		\begin{equation}
			Z_\alpha = \int_{\Omega_\alpha} \mathrm{d}\mathbf{x}\ e^{-\beta U(\mathbf{x})} \ .
			\label{eq:Z_molec}
		\end{equation}
		is the configuration integral of species $\alpha$ and
		\begin{equation}
			\Lambda = \prod_{i=1}^{3N} \lambda_i =  \prod_{i=1}^{3N} \sqrt{\frac{h^2}{2\pi m_i \kt}}
			\label{eq:deBroglie}
		\end{equation}
		is the product of de Broglie thermal wavelengths associated with atoms $i$.
		
		Because $\Omega_\mathrm{R}$ and $\Omega_\mathrm{P}$ comprise the entire configuration space available to the system, we have
		\begin{equation}
			Z = \int \mathrm{d}\mathbf{x}\ e^{-\beta U(\mathbf{x})} = Z_\mathrm{R} + Z_\mathrm{P} \ .
			\label{eq:Z_complete}
		\end{equation}
		Thus, the fraction of configuration space occupied by species $\alpha$ is the probability $\mathcal{P}(\alpha)$ that an observation of the system finds species $\alpha$:
		\begin{equation}
			\mathcal{P}(\alpha) = \frac{Z_\alpha}{Z}
			\label{eq:P_alpha}
		\end{equation}
		Every atomic arrangement can be described by a collective variable (CV), denoted by $\xi(\bx)$, which is, in general, a function of a subset of Cartesian coordinates. 
		The CV is chosen so that it has two non-overlapping domains that correlate with $\Omega_\mathrm{R}$ and $\Omega_\mathrm{P}$, as defined by the dividing surface. 
		We define the marginal probability density on the CV by
		\begin{equation}
			\rho(z) = \frac{1}{Z} \int \rmd \bx\ e^{-\beta U(\bx)}\ \delta\left[\xi(\bx) -z \right]
			= \left< \delta\left[\xi(\bx) -z \right] \right>
			\label{eq:rho_z}
		\end{equation}
		where the brackets denote the Boltzmann ensemble average over configuration space and $\delta$ is the Dirac delta function. 
		From \eqref{eq:rho_z} we deduce the relation
		\begin{equation}
			\int_{\Omega_\alpha} \rmd z\ \rho(z) =  \frac{Z_\alpha}{Z} = \mathcal{P}(\alpha) \ .
			\label{eq:P_alpha_rho}
		\end{equation}
		The Helmholtz free energy of species $\alpha$ is given by
		\begin{equation}
			F_\alpha = - \kt \ln q_\alpha
			\label{eq:F_alpha}
		\end{equation}
		Hence, using eqs.~(\ref{eq:Q_molec_integrated}), (\ref{eq:P_alpha_rho}), and (\ref{eq:F_alpha}) we express the reaction free energy as
		\begin{equation}
			\Delta F_\mathrm{RP} = F_\mathrm{P} - F_\mathrm{R} 
			= -\kt \ln \frac{Z_\mathrm{P}}{Z_\mathrm{R}}
			= - \kt \ln \frac{\mathcal{P}(\mathrm{P})}{\mathcal{P}(\mathrm{R})}
			\label{eq:F_reaction}
		\end{equation}
		It may be shown that the formula for the activation free energy is\cite{DietschreitDiestlerBombarelli2022}
		\begin{equation}
			\Delta F^\ddagger_\mathrm{RP} =
			- \kt \ln \frac{\rho(z_\mathrm{TS}) \left< \lambda_\xi \right>_{z_\mathrm{TS}}}{\mathcal{P}(\mathrm{R})} \ ,
			\label{eq:F_activation}
		\end{equation}
		where $z_\mathrm{TS}$ is the value of the CV at the dividing or transition state surface (TS) and
		\begin{equation}
			\lambda_\xi = \sqrt{\frac{h^2}{2 \pi \kt m_\xi}}
			\label{eq:lambda_xi}
		\end{equation}
		is the de Broglie thermal wavelength of the pseudo-particle associated with the CV. 
		The inverse effective mass of the pseudo-particle is given by
		\begin{equation}
			m_\xi^{-1} =  (\nabla_{{x}} \xi)^\mathrm{T} \mathbf{M}^{-1}(\nabla_{{x}} \xi) 
			\label{eq:m_xi} = | \nabla_\mwc \xi |^2 \ .
		\end{equation}
		$\nabla_\mwc$ signifies the gradient with respect to mass-weighted coordinates.
		(Note that the dimensions of $m_\xi^{-1}$ are dimensions of $\xi$ squared over mass times length squared.)
		In \eqref{eq:F_activation} $\left<\ \right>_{z_\mathrm{TS}}$ stands for the \enquote{$z$-conditioned} ensemble average over configuration space (i.e., the ensemble average under the constraint that the CV is fixed at $z_\mathrm{TS}$).
		
		\subsection{Reaction Entropy and Internal Energy}
		
		From \eqref{eq:F_reaction} we have the expression for the entropy of reaction:
		\begin{align}
			\Delta S_\mathrm{RP} & = - \frac{\partial \Delta F_\mathrm{RP}}{\partial T} 
			= k_\mathrm{B} \ln \frac{Z_\mathrm{P}}{Z_\mathrm{R}} + \kt \left[ \frac{\partial \ln Z_\mathrm{P}}{\partial T} -
			\frac{\partial \ln Z_\mathrm{R}}{\partial T} \right] \nonumber \\
			& = - \frac{\Delta F_\mathrm{RP}}{T} + \frac{1}{T} \left( \left< U \right>_\mathrm{P} - \left< U \right>_\mathrm{R} \right) \ ,
			\label{eq:S_reaction}
		\end{align}
		where the second line invokes eqs.~(\ref{eq:Z_molec}) and (\ref{eq:F_reaction}) and the subscript $\alpha$ on  $\left< U \right>_\alpha$ signifies the ensemble average over the configuration space of species $\alpha$.
		
		The internal energy of species $\alpha$ is given by
		\begin{align}
			E_\alpha & = k_\mathrm{B} T^2 \frac{\partial \ln q_\alpha}{\partial T} = \frac{\int_{\Omega_\alpha} \rmd \bx\ \int \rmd \bp\ e^{-\beta \mathcal{H}}\ \mathcal{H} }{\int_{\Omega_\alpha} \rmd \bx\ \int \rmd \bp\ e^{-\beta \mathcal{H}} } 
			\nonumber \\
			& = \frac{3}{2}N\kt + \left< U \right>_\alpha
			\label{eq:E_alpha}
		\end{align}
		where the second equality follows from \eqref{eq:Q_molec} and the second line from the fact that $\mathcal{H}$ is separable into a kinetic contribution depending only on the momenta and a configurational contribution depending only on coordinates (see \eqref{eq:Hamiltonian_full}).
		Using \eqref{eq:E_alpha}, we rewrite \eqref{eq:S_reaction} as
		\begin{equation}
			\Delta F_\mathrm{RP} = \Delta E_\mathrm{RP} - T\Delta S_\mathrm{RP} \ ,
			\label{eq:Helmholtz_reaction}
		\end{equation}
		where
		\begin{equation}
			\Delta E_\mathrm{RP} = E_\mathrm{P} - E_\mathrm{R} = \left< U \right>_\mathrm{P} - \left< U \right>_\mathrm{R}
			\label{eq:E_reaction}
		\end{equation}
		is the internal energy of reaction.
		We note the cancellation of the kinetic contributions.
		

		\subsection{Activation Entropy and Internal Energy}
		
		From \eqref{eq:F_activation} we have for the entropy of activation:
		\begin{align}
			\Delta S^\ddagger_\mathrm{RP} & = - \frac{\partial \Delta F^\ddagger_\mathrm{RP}}{\partial T} 
			\nonumber \\
			& = - \frac{1}{T} \left(\Delta F^\ddagger_\mathrm{RP} + \frac{\partial \ln \rho(z_\mathrm{TS})}{\partial \beta} \right. \nonumber \\
			& \qquad \qquad \left. + \frac{\partial \ln \left< \lambda_\xi \right>_{z_\mathrm{TS}}}{\partial \beta} - \frac{\partial \ln \mathcal{P}(\mathrm{R})}{\partial \beta} \right) \nonumber \\
			& = 
			\frac{1}{T} \left (- \Delta F^\ddagger_\mathrm{RP} + \frac{\left< U\ |\nabla_\mwc \xi|  \right>_{z_\mathrm{TS}}}{\left< |\nabla_\mwc \xi|  \right>_{z_\mathrm{TS}}} -  \frac{\kt}{2} - \left< U \right>_\mathrm{R} \right) \ ,
			\label{eq:S_act}
		\end{align}
		where the third line uses expressions for the partial derivatives derived in Appendix~A.
		
		It is instructive to rearrange \eqref{eq:S_act} to
		\begin{equation}
			\Delta F^\ddagger_\mathrm{RP} = \Delta E^\ddagger_\mathrm{RP} - T \Delta S^\ddagger_\mathrm{RP}
			\label{eq:Helmholtz_activation}
		\end{equation}
		where we define the activation internal energy by
		\begin{equation}
			\Delta E^\ddagger_\mathrm{RP} = \frac{\left< U\ |\nabla_\mwc \xi|  \right>_{z_\mathrm{TS}}}{\left< |\nabla_\mwc \xi|  \right>_{z_\mathrm{TS}}} -  \frac{\kt}{2} - \left< U \right>_\mathrm{R} 
			\label{eq:E_act}
		\end{equation}
		Through comparison with \eqref{eq:E_alpha} we infer that the internal energy at the TS is given by
		\begin{equation}
			E^\ddagger = \frac{3N -1}{2}\kt + \frac{\left< U\ |\nabla_\mwc \xi|  \right>_{z_\mathrm{TS}}}{\left< |\nabla_\mwc \xi|  \right>_{z_\mathrm{TS}}} \ .
			\label{eq:E_ddagger}
		\end{equation}
		%
		Note that the kinetic contribution is exactly missing the term corresponding CV being held at constant value. 
		The second term of \eqref{eq:E_ddagger} is interpreted as the mean configurational energy at the TS. 
		The expression for $\Delta E^\ddagger_\mathrm{RP}$ in \eqref{eq:E_act} can now be seen to be equivalent to $E^\ddagger - E_\mathrm{R}$, in which $(3N-1)/2$ kinetic terms cancel out.
		
		\subsection{Free-Energy Profile}
		
		In order to derive a gauge-invariant profile for the free energy, we now consider the system subject to the constraint that the CV $\xi(\mwc)$, a function of the (mass-weighted) coordinates, is fixed. 
		The free energy of the constrained system is now computed from the constrained partition function, which in turn is defined in terms of the constrained Hamiltonian. 
		Since the CV is in general a complicated function, it is convenient to transform to curvilinear coordinates, of which one is the CV  $q_1 = \xi(\mwc)$ and the remainder are orthogonal to $q_1$, i.e., the remainder satisfy the relation
		\begin{equation}
			(\nabla_\mwc q_1)^\mathrm{T}\ \nabla_\mwc q_i 
			= (\nabla_\mwc \xi)^\mathrm{T}\ \nabla_\mwc q_i 
			= \delta_{1i} |\nabla_\mwc \xi|^2 \ ,
			\label{eq:orthogonality_crit}
		\end{equation}
		where $\delta_{ij}$ is the Kronecker delta.
		Expressed in these curvilinear coordinates, the Hamiltonian assumes the form
		\begin{equation}
			\mathcal{H} = \frac{1}{2} p_1^2 |\nabla_\mwc \xi|^2 + \frac{1}{2} {\mathbf{p}_q'}^\mathrm{T} \mathbf{M}'^{-1} \mathbf{p}_q' + U(q_1, \mathbf{q}') \ ,
			\label{eq:Hamiltonian_curvilinear}
		\end{equation}
		where $\mathbf{p}_q' = (p_2, p_3, \dots, p_{3N})^\mathrm{T}$ 
		is the column vector of momenta conjugate to coordinates $\mathbf{q}' = (q_2, q_3, \dots, q_{3N})^\mathrm{T}$.
		
		We assume that the system is constrained so that the CV is fixed at $\xi(\mwc) = q_1 = z$ and $p_1 = 0$.
		Then the complete Hamiltonian in \eqref{eq:Hamiltonian_curvilinear} reduces to the constrained Hamiltonian
		\begin{equation}
			\mathcal{H}(z) =  \frac{1}{2} {\mathbf{p}_q'}^\mathrm{T} \mathbf{M}'^{-1} \mathbf{p}_q' + U(z, \mathbf{q}') \ .
			\label{eq:Hamiltonian_z} 
		\end{equation}
		The constrained partition function, defined by
		\begin{equation}
			Q(z) = \frac{1}{h^{3N-1}} \int \rmd \mathbf{q}'\ \int \rmd \mathbf{p}_q'\ e^{-\beta \mathcal{H}(z)}
			\label{eq:Q_z}
		\end{equation}
		can be expressed as
		\begin{equation}
			Q(z) = \frac{1}{h^{3N}} \sqrt{\frac{h^2}{2\pi \kt}} \int \rmd \mathbf{q}\ \int \rmd \mathbf{p}_q\ e^{-\beta \mathcal{H}}\ |\nabla_\mwc \xi|\ \delta[q_1 - z] \ .
			\label{eq:Q_z_expanded}
		\end{equation}
		That the relation in \eqref{eq:Q_z_expanded} is valid can be easily verified by substituting the expression given in \eqref{eq:Hamiltonian_curvilinear} and carrying out the integrations on $q_1$ and $p_1$.
		
		Transforming from curvilinear to Cartesian coordinates, we rewrite \eqref{eq:Q_z_expanded} as
		\begin{align}
			Q(z) & = Q \ \sqrt{\frac{h^2}{2\pi \kt}}\ \left< |\nabla_\mwc \xi|\ \delta[\xi(\bx) - z] \right> \nonumber \\
			& = Q\ \rho(z)\ \left< \lambda_\xi \right>_z
			\label{eq:Q_z_short}
			\ ,
		\end{align}
		where $Q = Z / \Lambda$ is implicitly defined as the partition function of the unconstrained system and the brackets indicate the average over the entire configuration space. 
		The second equality is obtained by using eqs.~(\ref{eq:rho_z}), (\ref{eq:lambda_xi}), and~(\ref{eq:m_xi}).
		
		We define the (Helmholtz) free-energy profile (FEP) by
		\begin{equation}
			F(z) = - \kt \ln Q(z) = F - \kt \ln \left[\rho(z)\ \left< \lambda_\xi \right>_{z} \right]
			\label{eq:F_z}
		\end{equation}
		where $F=-\kt \ln Q$ is the free energy of the unconstrained system. 
		We emphasize that $F(z)$ is the true free energy of the system upon which the condition $\xi(\bx)=z$ is imposed.
		The second term of the right side of \eqref{eq:F_z} can be interpreted as the work needed to constrain the system.
		
		From what is stated in the previous paragraph we conclude that the true free energy of the molecule constrained at the TS is given by
		\begin{align}
			F^\ddagger & = F(z_\mathrm{TS}) = - \kt \ln Q(z_\mathrm{TS}) \nonumber \\
			& = - \kt \ln \left[ Q\ \rho(z_\mathrm{TS})\ \left< \lambda_\xi \right>_{z_\mathrm{TS}} \right]
			\label{eq:Fddagger}
		\end{align}
		Using eqs.~(\ref{eq:Q_molec_integrated}) and~(\ref{eq:P_alpha_rho}), we can rewrite \eqref{eq:F_activation} as 
		\begin{align}
			\Delta F^\ddagger_\mathrm{RP} & = - \kt \ln \left[ \frac{Z\  \rho(z_\mathrm{TS})\ \left< \lambda_\xi \right>_{z_\mathrm{TS}} }{\Lambda\ q_\mathrm{R}} \right] \nonumber \\
			& = - \kt \ln \left[ Q\ \rho(z_\mathrm{TS})\ \left< \lambda_\xi \right>_{z_\mathrm{TS}} \right] + \kt \ln q_\mathrm{R} \nonumber \\
			& = F(z_\mathrm{TS}) - F_\mathrm{R} \ ,
		\end{align}
		where the third line follows from \eqref{eq:Fddagger}.
		Thus, as we expect, the activation free energy can be expressed as the difference between the free energy of the molecule constrained at the TS and the free energy of the reactant. 
		It is worth remarking that this formula takes the same form as the one derived by conventional transition state theory.\cite{Laidler1987}

		\begin{table*}[!bt]
			\centering
			\caption{A summary of the central formulas for free energy, internal energy, and entropy for reactions, activation, and profiles.
			}
			\begin{tabular}{r| c @{\hspace{5ex}} c @{\hspace{5ex}} c}
				& $F$ & $E$ & $S$ \\\hline
				Profile 
				& $F(z) = F - \kt \ln \left[\rho(z)\ \left< \lambda_\xi \right>_{z} \right]$ 
				& $E(z)  = \frac{3N - 1}{2}\kt + \frac{\left< U\ m_\xi^{-1/2} \right>_z}{\left< m_\xi^{-1/2} \right>_z}$
				& $S(z) = \frac{1}{T} \left( E(z) - F(z) \right)$ \\
				Reaction 
				& $\Delta F_\mathrm{RP} = - \kt \ln \frac{\mathcal{P}(\mathrm{P})}{\mathcal{P}(\mathrm{R})}$ 
				& $\Delta E_\mathrm{RP} = \left< U \right>_\mathrm{P} - \left< U \right>_\mathrm{R}$
				& $\Delta S_\mathrm{RP} = \frac{1}{T} \left( \Delta E_\mathrm{RP} - \Delta F_\mathrm{RP} \right)$ \\
				Activation 
				& $\Delta F^\ddagger_\mathrm{RP} = - \kt \ln \frac{\rho(z_\mathrm{TS}) \left< \lambda_\xi \right>_{z_\mathrm{TS}}}{\mathcal{P}(\mathrm{R})}$
				& $\Delta E^\ddagger_\mathrm{RP} = \frac{\left< U\ m_\xi^{-1/2} \right>_{z_\mathrm{TS}}}{\left< m_\xi^{-1/2} \right>_{z_\mathrm{TS}}} -  \frac{\kt}{2} - \left< U \right>_\mathrm{R} $
				& $\Delta S^\ddagger_\mathrm{RP}  = 
				\frac{1}{T} \left (\Delta E^\ddagger_\mathrm{RP} - \Delta F^\ddagger_\mathrm{RP}  \right)$ \\
			\end{tabular}
			\label{tab:summary}
		\end{table*}
		
		\subsection{Entropy and Internal-Energy Profiles}
		\label{sec:E_S_z}
		Using eqs.~(\ref{eq:Q_z_short}) and (\ref{eq:F_z}), we can express the entropy of the constrained system (i.e., the  entropy profile) as 
		\begin{align}
			S(z) & = - \frac{\partial F(z)}{\partial T} 
			= k_\mathrm{B} \ln Q(z) + \kt \frac{\partial }{\partial T} \ln Q(z) \nonumber \\
			& = -\frac{F(z)}{T} - \frac{1}{T}\frac{\partial}{\partial \beta}\left( \ln Q + \ln \rho(z) + \ln \left<\lambda_\xi\right>_z \right)\nonumber \\
			& = \frac{1}{T} \left( - F(z) + \frac{3N - 1}{2}\kt + \frac{\left< U\ |\nabla_\mwc \xi|  \right>_z}{\left< |\nabla_\mwc \xi|  \right>_z}\right) \ ,
			\label{eq:S_z}
		\end{align}
		where the last line was achieved by substitution of eqs.~(\ref{eq:dlnrhoz_dbeta}), (\ref{eq:dlambxi_dbeta}), and (\ref{eq:dlnQ_dbeta}) from Appendix~A.
		
		The  internal-energy profile (IEP) is
		\begin{align}
			E(z) & =  k_\mathrm{B} T^2 \frac{\partial \ln Q(z)}{\partial T} = - \frac{\partial \ln Q(z)}{\partial \beta} \nonumber \\
			& = \frac{3N - 1}{2}\kt + \frac{\left< U\ |\nabla_\mwc \xi|  \right>_z}{\left< |\nabla_\mwc \xi|  \right>_z}
			\label{eq:E_z}
		\end{align}
		where we again make use of eqs.~(\ref{eq:dlnrhoz_dbeta}), (\ref{eq:dlambxi_dbeta}), and (\ref{eq:dlnQ_dbeta}) to reach the second line.
		An alternative derivation of the relations in eqs.~(\ref{eq:S_z}) and (\ref{eq:E_z}) is given in Section~S2 of the Supporting Information (SI).
		
		Comparing \eqref{eq:E_z} with \eqref{eq:E_ddagger}, we see that $E^\ddagger = E(z_\mathrm{TS})$.
		We interpret the expression in \eqref{eq:E_z} as the internal-energy profile of the constrained system. 
		Thus, the internal energy of the constrained molecule is the sum of the kinetic energy (independent of $z$) and the constrained configuration energy (dependent on $z$).
		Because one degree of freedom is fixed by the constraint, the kinetic energy of the constrained system is $\kt/2$ less than that of the unconstrained system.
		The factor $|\nabla_\mwc \xi|$ stems from the transformation from mass weighted to curvilinear coordinates. 
		The magnitude of the gradient is a measure of the local spacing between isosurfaces of constant $z$ and therefore of the variation in the $(3N -1)$-dimensional subspace of configuration space on which the $z$-conditioned ensemble average is computed.
		As shown in Appendix~B, this form of the weighted average commonly results when the constrained ensemble average is computed in the phase space of the unconstrained system. 
		We note that if the CV is linear in the Cartesian coordinates, then $|\nabla_\mwc \xi|$ is constant and $E(z)$ reduces to the  form $E(z) = \frac{3N - 1}{2}\kt + \left< U\right>_z$. 
		This is because the successive isosurfaces are equidistant.
		
		Combining eqs.~(\ref{eq:S_z}) and~(\ref{eq:E_z}) yields the following relation between the several constrained profiles:
		\begin{equation}
			F(z) = E(z) - TS(z)
			\label{eq:Helmholtz_z}
		\end{equation}
		For convenience we gather the various expressions thus far derived in Tab.~\ref{tab:summary}.
		
		Because of the impracticability of computing $F$ (the free energy of the unconstrained system), we compute the profiles relative to a zero point $z_0$, where we set the free energy $F(z_0)$ to zero. 
		That is, the profile consists of a plot of  $F(z) - F(z_0)$. 
		The same is true of the other profiles.

		\subsection{Constrained Mean Force}
		
		Differentiating the first of the two equalities of \eqref{eq:F_z} with respect to $z$ (the value of the CV)  gives
		%
		%
		\begin{align}
			- \frac{\partial F(z)}{\partial z} 
			& = \frac{ \kt \int \rmd \bq' \int \rmd \bp_q'\ e^{-\beta \mathcal{H}(z)}\ \frac{-\beta \partial \mathcal{H}(z)}{\partial z}}{\int \rmd \bq' \int \rmd \bp_q'\ e^{-\beta \mathcal{H}(z)}} \nonumber \\
			& = \left< - \frac{\partial \mathcal{H}(z)}{\partial z} \right>_{q_1=z, p_1=0} 
			\label{eq:dF_dz_meanforce} \\
			& = \left<- \frac{\partial U}{\partial z}  - \frac{\partial K}{\partial z}\right>_{q_1=z, p_1=0}  \nonumber
		\end{align}
		In the last line $K$ and $U$ stand for the kinetic and potential energy contributions to $\mathcal{H}(z)$.
		In general we expect the mass matrix $\mathbf{M}'$  to depend on the curvilinear coordinates. 
		The subscripted brackets  $\left< \ \right>_{q_1=z, p_1=0}$ signify the ensemble average over the curvilinear phase space, excluding $q_1=z$ and $p_1=0$.
		
		We next observe that 
		$- \frac{\partial \mathcal{H}(z)}{\partial z}$ is the instantaneous force on the CV for any fixed phase $\bq,\bp_q$. 
		In other words, the opposite force is needed to maintain the constraint. 
		Hence, the right side of \eqref{eq:dF_dz_meanforce} is the mean force on the CV averaged over the complementary phase space (i.e., the whole phase space exclusive of  $q_1,p_1$). 
		Thus, we may regard $F(z)$ as the potential of mean constrained force on the CV. 
		
		An alternative approach, which is detailed in Sections~S4 and S5 of the SI, yields the following expression:
		\begin{widetext}
			\begin{align}
				- \frac{\partial F(z)}{\partial z} 
				& = \frac{1}{\left< |\nabla_\mwc \xi |\right>_z} \left<  |\nabla_{\mwc} \xi|\ \left( - \nabla_\mwc U \cdot \frac{\nabla_\mwc \xi  }{|\nabla_\mwc \xi|^2} + \frac{\kt}{|\nabla_\mwc \xi|} \left[ \nabla_\mwc \cdot \frac{\nabla_\mwc \xi}{|\nabla_\mwc \xi|} \right] \right)
				\right>_z 
				\label{eq:dFz_dz}
			\end{align}        
		\end{widetext}
		%
		Note that this expression includes only $z$-conditioned averages for the unconstrained system.
		Though the expressions given by eqs.~(\ref{eq:dF_dz_meanforce}) and (\ref{eq:dFz_dz}) have very different form, a correlation can be made. 
		The first term of \eqref{eq:dFz_dz} corresponds to the contribution from the PES; the second term, which possesses a \enquote{geometric} character, arises from the kinetic energy. 
		The form of \eqref{eq:dFz_dz}, the ensemble average weighted by the absolute of the gradient, is analogous to the weighted average of the potential energy in the IEP and commonly appears when constrained ensemble averages are computed in the phase space of the unconstrained system\cite{Fixman1974, Carter1989} (see \eqref{eq:Fixman} of Appendix~B).

		\subsection{Relation Between Free-Energy Profiles for CVs With Parallel Gradients}
		
		
		In this Section we consider two CVs, where we take $\xi(\mwc)=q_1$ to be one CV and the other CV to be a function of only $q_1$, i.e., we have
		\begin{equation}
			\phi(\mwc) = f(q_1)
			\label{eq:phi_def}
		\end{equation}
		where we assume that $f^{-1}$ exists. 
		
		Taking the gradient of \eqref{eq:phi_def}, we get
		\begin{equation}
			\nabla_\mwc \phi = f'(q_1) \nabla_\mwc q_1 = f'(q_1) \nabla_\mwc \xi
		\end{equation}
		Thus, since the gradient of $\phi$ is proportional by the derivative of $f$ to the gradient of $\xi$, the gradients of the two CVs are parallel everywhere in configuration space.
		
		We assume that $q_1$, together with $\mathbf{q}'$ (the subset of the remaining $3N-1$ coordinates), forms a complete set $\{q_1, \mathbf{q}' \}$ of mutually orthogonal coordinates. 
		Expressed in mass-weighted coordinates, they obey the condition
		\begin{equation}
			(\nabla_\mwc q_i)^\mathrm{T}\ \nabla_\mwc q_j
			= \delta_{ij} |\nabla_\mwc q_i|^2
			\label{eq:orthogonality_complete}
		\end{equation}
		(Note that this is a stronger condition than the one we impose in the derivation in Section~IID (see \eqref{eq:orthogonality_crit}).
		Then, the gradient of $\phi$ is orthogonal to the gradients of the curvilinear coordinates $\{\mathbf{q}'\}$, and therefore $\{ \phi, \mathbf{q}'\}$ also constitutes a complete set of orthogonal curvilinear coordinates.
		
		As shown in the Section~S3 of the SI two such CVs satisfy the following relationship:
		\begin{equation}
			Q^\xi(z) = Q^\phi(f(z)) = Q^\phi(z') \ ,
		\end{equation}
		with $z' = f(z)$.
		The corresponding relation between the FEPs is, according to the definition in \eqref{eq:F_z},
		\begin{equation}
			F^\xi(z) = F^\phi(f(z)) = F^\phi(z')
			\label{eq:F_parallel_CVs}
		\end{equation}
		%
		Thus, both FEPs assume the same value on the same hypersurface in configuration space.
		Hence, the FEP is gauge-invariant.

		\section{Connection With Prior Work}
		
		\subsection{Comparison of the FEP with PMF}
		
		The PMF is defined as a function of the marginal probability density on the CV (\eqref{eq:rho_z}) by\cite{Kirkwood1935, Kumar1992, Roux1995}
		\begin{equation}
			A(z) = - k_\mathrm{B}T \ln \rho(z)
			\label{eq:A_z}
		\end{equation}
		To appreciate the rather remarkable relation in \eqref{eq:F_parallel_CVs} between FEPs whose CVs have parallel gradients, namely that the FEP of one as a function of its CV is equal to the FEP of the other as a function of its CV, we compare it with the analogous relation that obtains for the PMF. 
		A detailed derivation in Section~S3 of the SI gives
		\begin{equation}
			A^\xi(z) = A^\phi(f(z)) - \kt \ln \left|f'(z) \right|
		\end{equation}
		Thus, the PMF does not obey a relation analogous to that in \eqref{eq:F_parallel_CVs}. 
		It is possible that two CVs having parallel gradients assume different values on the same hypersurface (in configuration space). 
		Hence, the difference between the values of the PMF at two different points on the CV cannot be a true free-energy difference.

		The derivative of the PMF is usually given as\cite{Wong2012, Comer2015}
		\begin{align}
			- \frac{\partial A(z)}{\partial z} & = \left< - \nabla U \cdot \frac{\nabla \xi  }{|\nabla \xi|^2} + {\kt} \left[ \nabla \cdot \frac{\nabla \xi}{|\nabla \xi|^2} \right]
			\right>_z 
			\label{eq:dAz_dz}
		\end{align}
		The first term of \eqref{eq:dAz_dz} is the contribution to the mean force from the PES. 
		The second term is often interpreted to be of a purely geometric character, though it can be argued that it is the kinetic contribution to the mean force, as it stems from the derivative of the kinetic energy in curvilinear coordinates (see Section~S6 of the SI for a detailed derivation of \eqref{eq:dAz_dz}). 
		However, note the difference in this second term between \eqref{eq:dAz_dz} and \eqref{eq:dFz_dz}.


		\subsection{Comparison with the Geometric Free-Energy Profile}
		
		An alternative  free-energy profile has been proposed, namely the \enquote{geometric} free-energy profile (GFEP), one version of which is given by\cite{ hartmann2007, hartmann2011two}
		\begin{equation}
			A^G(z) = -\kt \ln \left[ \rho(z) \left<|\nabla \xi|\right>_z \right]
			\label{eq:A_geometric}
		\end{equation}
		Some definitions\cite{Vanden2005, Bal2020} incorporate an arbitrary constant in the argument of the logarithm in order to render the argument dimensionless. 
		From an examination of the various publications dealing with the GFEP it is not evident whether the gradient of the CV is with respect to the Cartesian or mass-weighted coordinates.\cite{Bal2020} 
		If we assume that the gradient is the mass-weighted one, then we find that the GFEP differs from the FEP only by a constant. 
		
		Previous studies\cite{hartmann2011two} have shown that the GFEP satisfies the relation in \eqref{eq:F_parallel_CVs}. 
		However, our derivation of the FEP clearly reveals the physical constants (contained in the de Broglie wavelength $\lambda_\xi$) missing from the GFEP. 
		Further, no arbitrary constant needs be introduced. 
		Moreover, our derivation gives the GFEP a clear physical interpretation, which has been lacking until now.\cite{hartmann2011two, Bal2020}


		\section{APPLICATIONS}
		
		We turn next to applications of the formulas in Tab.~I to systems of increasing complexity. 
		
		\subsection{Dissociation of Contact Pair}
		
		We consider the dissociation of a contact pair to a solvent-separated pair of atoms in solution. 
		We assume that the PES depends only on the distance between the atoms of the pair. 
		We further assume that the solvent is not dynamically involved in the process (i.e., solvent molecules rearrange rapidly on the time scale of the process).
		The CV is taken to be the distance between the atoms. 
		A detailed treatment is presented in Section~S7 of the SI. 
		
		The FEP is given by
		\begin{equation}
			F(z) = U(z) - \kt \ln \frac{4\pi z^2}{\lambda^2} \ ,
			\label{eq:cp_Fz}
		\end{equation}
		where $\lambda$ is the thermal de Broglie wavelength. 
		The entropy profile is 
		\begin{equation}
			S(z) = \kb \left( 1 +  \ln \frac{4\pi z^2}{\lambda^2} \right)
			\label{eq:cp_Sz}
		\end{equation}
		Combining eqs.~(\ref{eq:cp_Fz}) and~(\ref{eq:cp_Sz}), we obtain the internal-energy profile
		\begin{equation}
			E(z) = U(z) + \kt 
			\label{eq:cp_Ez}
		\end{equation}
		as the sum of the configurational and kinetic energies (each of the two spherical angular terms of the kinetic energy contributes $\kt / 2$).
		
		Regarding $F(z)$ as the potential of mean constrained force (see Section~IIF), we obtain from \eqref{eq:cp_Fz}  the mean force acting on the CV 
		\begin{align}
			-\frac{\partial F(z)}{\partial z} & = -\frac{\partial E(z)}{\partial z} + T \frac{\partial S(z)}{\partial z} \nonumber \\
			& = - \frac{\partial U(z)}{\partial z} + \frac{2\kt}{z}
		\end{align}
		The force apparently comprises two components, one arising from the energetic contribution to the CFEP and the other from the entropic contribution. 
		It is noteworthy that the entropic component falls off extremely slowly with increasing $z$, pushing the atoms of the contact pair apart. 
		In contrast, the energetic component decreases relatively rapidly with increasing separation. 
		The derivation of constrained internal energy and entropy profiles makes it possible to identify the thermodynamic origins of contributions to the mean force.
		
		\begin{figure}[bt]
			\centering
			\includegraphics[width=\linewidth]{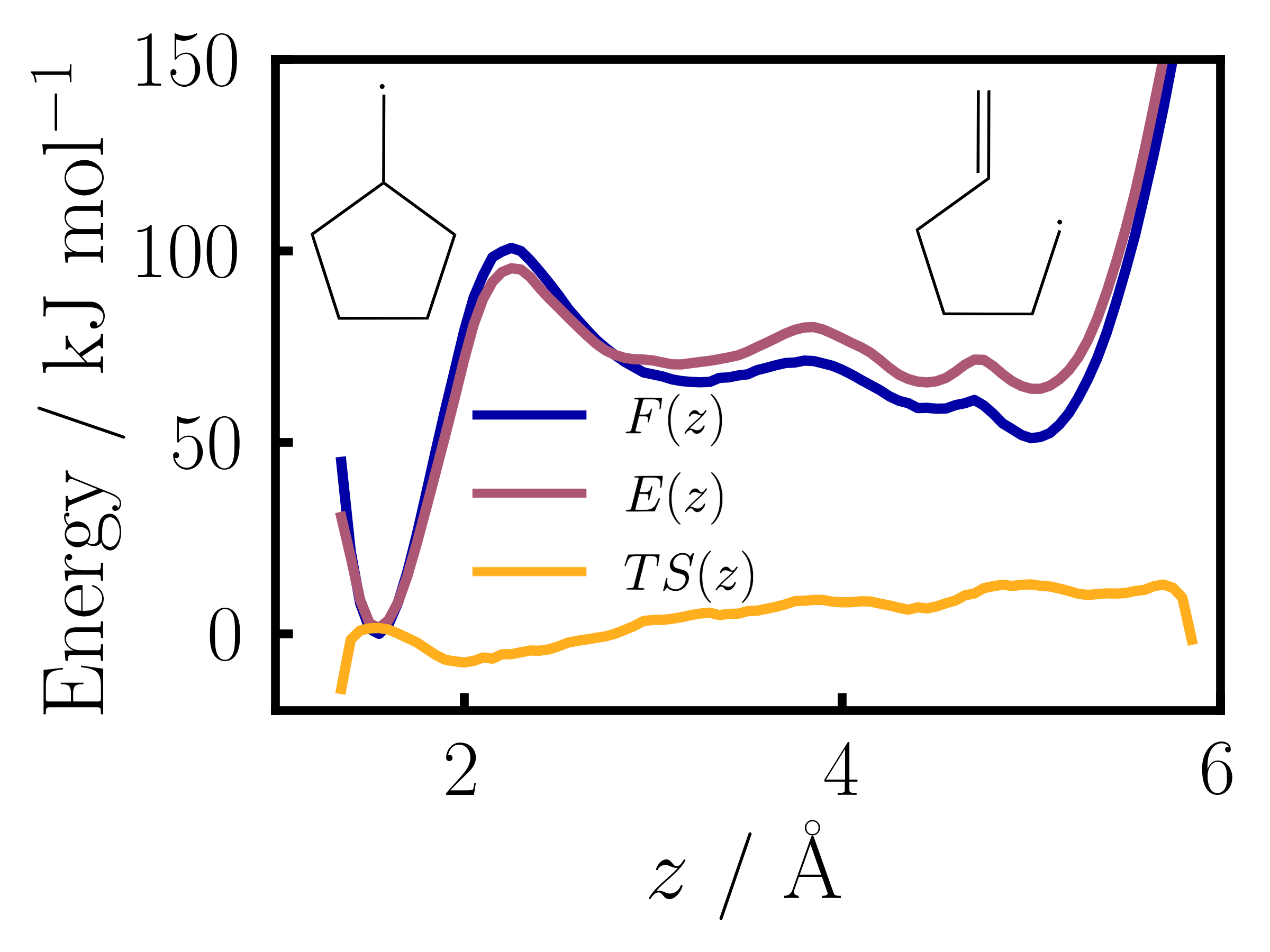}
			\caption{Free-energy, internal-enrgy and entropy profiles for the intramolecular cyclization of the 5-hexenyl radical.}
			\label{fig:cfep_cyclization}
		\end{figure}

		\subsection{Intramolecular Radical Cyclization}
		
		Here we revisit the intramolecular cyclization of the 5-hexenyl radical reported in our previous work\cite{DietschreitDiestlerBombarelli2022}.
		Fig.~\ref{fig:cfep_cyclization} displays Lewis structures for reactant (R) and product (P). 
		The process of cyclization involves the formation of a new C-C single bond accompanied by the conversion of a C-C double bond to a single bond. 
		The natural choice of CV is the distance between the C1 and C5 carbon atoms: $\xi = \left|\mathbf{r}_\mathrm{C5} - \mathbf{r}_\mathrm{C1} \right|$. 
		We simulated the system by the WTM-eABF\cite{lesage2017smoothed, fu2018zooming, fu2019taming} algorithm at 300~K using \textit{ab initio}  MD  (see Ref.~\citenum{DietschreitDiestlerBombarelli2022} for details). 
		The MD frames were unbiased with MBAR.\cite{Hulm2022}
		
		\begin{table}[bt]
			\centering
			\caption{Reaction and activation energies and entropies for forward and reverse cyclization reactions.}
			\begin{tabular}{r| r r r}
				& $\frac{F}{\mathrm{kJ\ /\ mol}}$ & $\frac{E}{\mathrm{kJ\ /\ mol}}$ & $\frac{S}{\mathrm{J\ /\ mol\ K}}$ \\\hline
				Reaction (R$\rightarrow$P) & 
				-49.1 & -64.0 & -49.9 \\
				Activation (R$\rightarrow$P) &  
				48.2 & 29.2 & -63.3 \\
				Activation (P$\rightarrow$R) &  
				97.3 & 93.3 & -13.4
			\end{tabular}
			\label{tab:cyclization}
		\end{table}
		
		Fig.~\ref{fig:cfep_cyclization} shows plots of the profiles. 
		The entropy profile is calculated as the difference between the free-energy and internal-energy profiles. 
		We take the reference point to be $z_\mathrm{TS} = 2.2$~\AA\  (i.e., the transition state). 
		All configurations with $z>2.2$~\AA\  belong to R; configurations with $z<2.2$~\AA\ belong to P. 
		It is apparent that the FEP and internal-energy profile (IEP) are almost equal for P. 
		At the transition state the FEP assumes a slightly greater value than the IEP, on account of the low entropic contribution. 
		Since R is an open chain, it has a much higher entropy than the closed cyclic P. 
		Moreover, the entropy of R increases as the chain unfolds, although at large $\xi(\mathbf{x})$ the chain becomes stretched and stiff and the entropy decreases sharply.

		In Tab.~\ref{tab:cyclization} the reaction and activation energies and entropies are listed. 
		These are in excellent agreement with conclusions drawn from the profiles. 
		The gain in free energy by the cyclization is 15~kJ/mol smaller than the reaction internal energy because of the entropy difference between open chain and cyclic molecule.
		Tab.~\ref{tab:cyclization} also shows that the activation free and internal energies are almost identical for the ring opening but differ by almost 20~kJ/mol for the ring closure. 
		The latter discrepancy is due to the strongly increased entropy of the open chain, which explains the very negative activation entropy for the ring closure.

		
		\begin{figure*}[tb]
			\centering
			\includegraphics[width=\linewidth]{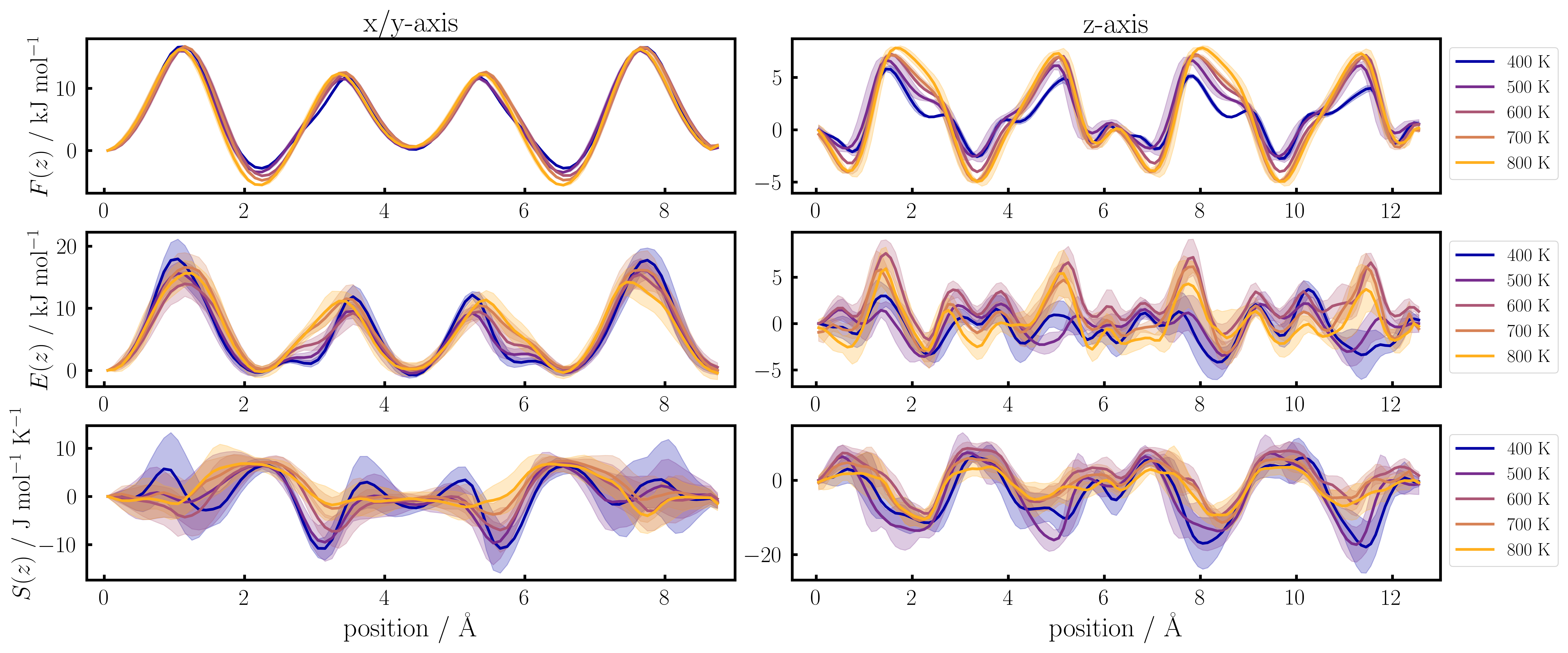}
			\caption{Free- and internal-energy and entropy profiles along unit cell axes of LGPS. 
				The estimated error (indicated by lightly shaded region) is taken to be the standard deviation, computed as described in Secotion~S9 of the SI.}
			\label{fig:LGPS_profiles}
		\end{figure*}
		
		\subsection{Mobility of Li$^+$ Ions in a Solid Electrolyte}
		
		LGPS (empirical formula  Li$_{10}$Ge(PS$_6$)$_2$) is a solid electrolyte with high concentration of mobile cations (Li$^+$). 
		Its high ionic conductivity is comparable with that of commercial organic liquid electrolytes used in lithium-ion batteries.\cite{Kamaya2011AConductor, Kato2020Li10GeP2S12-TypeTransportation} 
		LGPS has a BCC-like lattice with alternating  PS$_4$- and GeS$_4$-tetrahedra. 
		As the  Li$^+$-sublattice is fully occupied, there is substantial Coulomb repulsion between neighboring cations, which is understood to result in concerted hopping of ions and consequently the high ionic conductivity.\cite{He2017OriginConductors, Xu2012One-dimensionalConductor} 
		
		Here we employ simulation data from Ref.~\citenum{Gavin2023}  to compute the FEP and entropy profile along the Cartesian axes aligned with the cuboid unit cell. 
		Ref.~\citenum{Gavin2023} performed MD simulations of 4x4x4 supercells at temperatures ranging from 400~K to 800~K.  
		We take the CV to be the Cartesian component of the position of a given Li$^+$ relative to the origin of the unit cell in which it is found in a given MD frame. 
		Since all Li$^+$ ions are equivalent, we compute the profiles as appropriate averages over each ion
		(see Section~S9 of the SI for a detailed description). 
		Figure~S1 of the SI shows orthographic plots of the MD frames. 
		The conductive channels along the z-axis are clearly distinguished from the inhibited diffusion in the x-y-plane at 400~K.
		In contrast, at 800~K diffusion generally increases and occurs across all axes.   
		Because of the symmetry in LGPS, the profiles in the x- and y-directions are identical. 
		
		Fig.~\ref{fig:LGPS_profiles} displays plots of the IEP, FEP, and entropy profile for a range of temperatures. 
		The minima in the IEP at low temperature indicate sites that can be occupied by Li$^+$, corresponding to known interstices in the LGPS crystal structures.\cite{Bhandari2016} 
		Many of these minima disappear at higher temperature because  LGPS undergoes a phase transition that shuts off transport in the x and y direction\cite{Kato2020Li10GeP2S12-TypeTransportation, Weber2016StructuralLi10GeP2S12, Miwa2021MolecularPotential, Gavin2023} 
		Because channels in the the x- and y-directions are not aligned, the structure of the profiles along the z-axis is noticeably more complicated. 
		The entropy profiles along all axes share the trend that with increasing temperature high-entropy regions broaden and low-entropy regions flatten. 
		Hence, the FEPs become significantly smoother with increasing temperature. 
		
		The deep minima in the FEPs for the x- and y-directions correspond to the ion channels through which conduction along the z-direction occurs. 
		The smaller range of the FEP along the z-axis suggests lower activation barriers for ion motion in this direction.
		The FEPs along the x/y-axis change little from 400~K to 800~K which is in agreement with the markedly increased Li$^+$ motion along these axes with increased temperature; the activation free energy for hops between minima changes little, though the system temperature is doubled.

		\subsection{Mobility of  Copper-Ammonia Complex in a Nanoporous Catalyst}

		\begin{figure*}[!hbt]
			\centering
			\includegraphics[width=\linewidth]{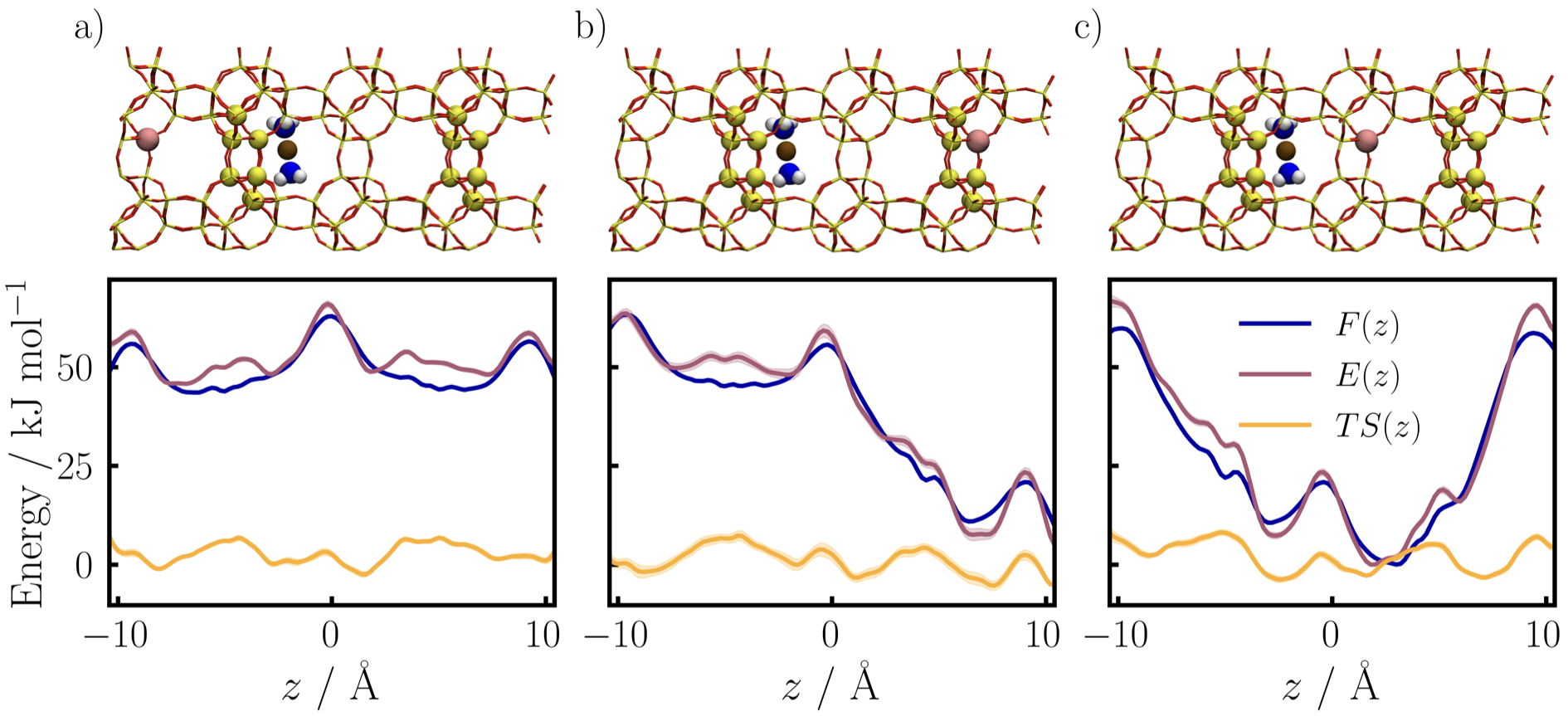}\\
			\caption{Upper panel: Depictions of the three systems, showing the position of the aluminum atom (salmon) relative to the two cavities between which the copper complex (Cu (brown), N (blue), H (white)) migrates.
				Oxygens are shown in red. The axis of diffusion is taken to be the vector between the centers of the outer two 8-rings, whose Si atoms are shown as yellow spheres.
				Lower panel: FEP, IEP, and entropy profile for the three systems depicted directly below. 
				a) the Al atom one cell removed, b) the Al atom located in one of the outer rings, and c) Al atom in the central 8-ring.
				The lowest value of FEP and IEP in c) have been set to zero. The profiles in a) and b) have been shifted along the abscissa to reflect the translational symmetry between the systems. 
				The CV value at the dividing surface between the cavities is around $z=0$ for each system.
				The error, shown as shaded area about each profile, is the standard deviation, determined as described in Section~S10 of the SI.
			} 
			\label{fig:CFEPChab}
		\end{figure*}

		Zeolites are nanoporous aluminosilicates used for catalysts and separations. 
		In particular copper exchanged materials are used in the selective catalytic reduction of nitrogen oxides in diesel engines.\cite{KWAK2010187,Gao2013, Nuria2015, Borfecchia2018, PEDEN2019} 
		We study the migration of the [Cu(NH$_3)_2]^+$-complex between two cavities in chabazite (Fig.~\ref{fig:CFEPChab}, upper panel). 
		The migration reaction is the movement of the Cu-complex from one cavity to the other through the central 8-ring (8 silicon sites). 
		The simulation cell (a 4x2x2 supercell) contains a single aluminum atom, which is placed at the position of an Si in an 8-ring.
		It introduces a negatively charged site and therefore neutralizes the charge and strongly attracts the Cu-complex. 
		As the axis of diffusion we take the vector connecting the two centers of mass of the outer 8-rings (highlighted as yellow spheres in Fig.~\ref{fig:CFEPChab}). 
		The CV is the projection of the vector connecting the midpoint between the centers of mass of the two outer 8-rings and the center of mass of the Cu-complex onto the axis of diffusion. 
		Thus, the zero point of the CV is roughly at the transition state, which is located at the central 8-ring.
		The three systems depicted in Fig.~\ref{fig:CFEPChab} differ by the Al position, which is translated by one or two cavities, respectively.
		The relative position of the aluminum has important consequences in the driving force and the barrier for diffusion.
		
		\begin{table}
			\centering
			\caption{Reaction and activation free- and internal energies and entropies for migration of the Cu complex in chabazite. 
				Energy in units of kJ/mol and entropy in units of J/mol~K. 
			}
			\label{tab:chab_values}
			{
				\begin{tabular}{r| r r | r r | r r}
					\multicolumn{1}{c}{} & \multicolumn{2}{c}{Fig.~\ref{fig:CFEPChab}a} & \multicolumn{2}{c}{Fig.~\ref{fig:CFEPChab}b} & \multicolumn{2}{c}{Fig.~\ref{fig:CFEPChab}c} \\
					$X$ & $\Delta X$ & $\Delta X^\ddagger$ & $\Delta X$ & $\Delta X^\ddagger$ & $\Delta X$ & $\Delta X^\ddagger$ \\
					$F$ & $0.6$ & $32.2$  & $-31.0$ & $24.1$  & $-10.9$ & $20.2$ \\
					$E$ & $1.8$ & $18.9$  & $-41.5$ & $8.2$   & $-7.1$  & $14.5$ \\
					$S$ & $3.0$ & $-31.4$ & $-24.7$ & $-37.5$ & $9.1$   & $-13.4$
				\end{tabular}
			}
		\end{table}

		80-ns WTM-eABF simulations were performed at 423~K for each Al position using a neural force field~\cite{Millan2023} enhancing the movement of the Cu-complex along the CV. 
		The profiles are displayed in Fig.~\ref{fig:CFEPChab}.
		In every system, as the Cu-complex approaches the 8-ring ($z$ values around to -10, 0, and +10), it comes in contact with the scaffold and the ammonia ligands form hydrogen bonds with the oxygens of the scaffold, thereby lowering the energy and entropy (see Fig.~S2 of the SI for an analysis of this effect). 
		This results in internal energy and entropy minima adjacent to an 8-ring. 
		Away from the 8-ring, within the cage, the complex is free to move and has higher entropy. 
		At the same time the stabilization through H-bonds between the complex and the scaffold can be formed only in directions normal to the CV.
		Hence, the IEP is elevated within the cages.

		We expect the profiles to reflect the translational symmetry introduced by the Al position, i.e., the right half of the profiles in Fig.~\ref{fig:CFEPChab}a) should be identical to the left half of those in Fig.~\ref{fig:CFEPChab}b). 
		The same applies to respective halves of Fig.~\ref{fig:CFEPChab}b) and c). 
		However, the WTM-eABF simulations do not enhance the sampling orthogonal to the CV, and therefore the diffusion of the Cu-complex normal to the CV is a \enquote{rare event}.
		The FEP almost perfectly fulfills the translational symmetry.
		However, deviations between the IEPs are visible. 
		
		The profiles in Fig.~\ref{fig:CFEPChab}a) are the most symmetrical because the Al is one-cell removed, so its effect on reactant and product cavities is nearly similar. 
		The central barrier of the FEP is higher than the outer two barriers, as it is farthest from the Al. 
		The large barriers in the FEP coincide with those of the IEP. 
		
		In sharp contrast with Fig.~\ref{fig:CFEPChab}a), Fig.~\ref{fig:CFEPChab}b) exhibits a large asymmetry between the profiles in the R and P cavities, reflected in the large reaction internal and free energies (see Tab.~\ref{tab:chab_values}). 
		This is a consequence of the strong attraction between the negatively charged site, now located in one of the outer rings, and the positively charged Cu-complex. 
		Since the 8-ring separating the cavities is not planar, but has a crown-like structure, the Al is more accessible in one cavity than in the other. 
		Hence, the asymmetry persists in the profiles of Fig.~\ref{fig:CFEPChab}c), which pertain to the situation where the Al is located in the central 8-ring. 
		This in agreement with previous studies.\cite{Millan2021}
		
		Tab.~\ref{tab:chab_values} shows that for the two almost symmetrical systems (aluminium far removed and aluminum at the center) the activation internal energy differs by only 4~kJ/mol, whereas the activation free energy for diffusion from left to right differs by 12~kJ/mol.
		The strong attraction between Cu-complex and anion-site causes a general lowering in entropy and consequently a lowering of the activation entropy. 
		In contrast, the activation entropy for the forward reaction is very similar for the two systems, where the aluminum is not located in the ring through which diffusion occurs.

		\section{Conclusion}
		
		The main contribution of this article is a systematic approach to the description of reactive systems through the concept of the \enquote{profile}. 
		The profile of a given property is a function of only the CV. 
		It is generated by computing the property of the system under the constraint that the CV is fixed. 
		The profiles of central interest are those of the free energy, internal energy and entropy. 
		We demonstrate the physical importance of the profiles through their connection with activation energies and entropies. 
		Additionally, we show how these profiles can be derived from a meaningful partition function.
		
		Importantly, contrary to what their association with a constrained Hamiltonian would suggest, all three profiles can easily be obtained from the much more commonly performed unconstrained simulations without the need to freeze the motion along the CV.
		All important equations are formulated as averages over the unconstrained configuration space.
		
		
		The application of the theory to dissociation of a contact pair demonstrates the enhanced interpretability of the contributions to the mean forces in terms of internal energy and entropy.
		The applications to chemically realistic models highlight the utility of the two additional profiles and how they can guide chemical understanding of reactions.
		We hope that other researchers find the expressions developed in this work useful.

		\section*{Supplementary Information}
		
		The SI contains: i) alternative derivations of the constrained internal energy and entropy profiles; ii) derivation of relations between FEPs and PMFs corresponding to different CVs with parallel gradients; iii) a justification for the usage of the pseudo inverse,; iv) derivation of the mean forces given in eqs.~(44) and~(51); v) a  proof that the PMF is the potential of mean force; vi) a detailed treatment of the contact pair; vii) computational details for the three realistic chemical systems.

		\begin{acknowledgements}
			We thank Reisel Millan, Gavin Winter, and Andreas Hulm for providing access to their force fields and simulation data, as well as for stimulating conversations regarding the chemical systems.
			J.C.B.D. is thankful for the support of the Leopoldina Fellowship Program, German National Academy of Sciences Leopoldina, grant number LPDS 2021-08.
			R.G.-B. acknowledges support from the Jeffrey Cheah Career Development Chair.
		\end{acknowledgements}
		
		\section*{Author Declarations}
		\subsection*{Conflict of Interest}
		The authors have no conflicts of interest to disclose.

		\section*{Data Availability Statement}
		
		MD output files and Jupyter notebooks used for the analysis can be found at: \url{https://doi.org/10.5281/zenodo.7809206}
	
	\section*{References}
	\bibliography{references}

	
	\appendix
	\section{Partial Derivatives}
	\label{App:PartialDerivatives}
	We begin with
	\begin{align}
		\frac{\partial \ln \rho(z)}{\partial \beta} & = 
		\frac{1}{\rho(z)} \frac{\partial}{\partial \beta}\left( \frac{1}{Z} \int \rmd \bx\ e^{-\beta U(\bx)}\ \delta[\xi(\bx) - z]\ \right) \nonumber \\
		& =  \frac{-1}{\rho(z)}  \left( \frac{1}{Z^2}\frac{\partial Z}{\partial \beta} \int \rmd \bx\ e^{-\beta U(\bx)}\ \delta[\xi(\bx) - z]\ \right. \nonumber \\
		& \qquad \qquad \left. + \frac{1}{Z}\int \rmd \bx\ e^{-\beta U(\bx)}\  U(\bx)\ \delta[\xi(\bx) - z]\ \right) \ .
		\label{eq:dlnrhoz_dbeta_beginning}
	\end{align}
	Using the relation
	\begin{equation}
		\frac{\partial Z}{\partial \beta} = - Z \left< U \right>
		\label{eq:dZ_dbeta}
	\end{equation}
	we rewrite \eqref{eq:dlnrhoz_dbeta_beginning} as
	\begin{align}
		\frac{\partial \ln \rho(z)}{\partial \beta} & = 
		\frac{1}{\rho(z)}  \left( \left< U \right> \rho(z) - \left< U(\bx)\ \delta[\xi(\bx) - z]\right> \right) \nonumber \\
		& =  \left< U \right> -  \left< U \right>_z \ .
		\label{eq:dlnrhoz_dbeta}
	\end{align}
	From \eqref{eq:lambda_xi} we obtain
	\begin{align}
		\frac{\partial \ln \left< \lambda_\xi \right>_z}{\partial \beta} & = 
		\frac{\partial}{\partial \beta} \left( \ln\sqrt{h^2 \beta/2\pi} + \ln \left< |\nabla_\mwc \xi|  \right>_z \right) \nonumber \\
		& = \frac{\kt}{2} + \frac{1}{\left< |\nabla_\mwc \xi|  \right>_z} \frac{\partial \left< |\nabla_\mwc \xi|  \right>_z}{\partial \beta} \ .
		\label{eq:dlambxi_dbeta_beginning}
	\end{align}
	According to the definition of the $z$-conditioned ensemble average, we have
	\begin{align}
		\left< |\nabla_\mwc \xi| \right>_z  & = 
		\frac{\left<|\nabla_\mwc \xi|\ \delta[\xi(\bx) - z]\right> }{\left< \delta[\xi(\bx) - z]\right>} \nonumber \\
		& = \frac{\int \rmd \bx\ e^{-\beta U(\bx)}\ |\nabla_\mwc \xi|\ \delta[\xi(\bx) - z]}{\int \rmd \bx\ e^{-\beta U(\bx)}\  \delta[\xi(\bx) - z]} \ . 
		\label{eq:invmxi_zave}
	\end{align}
	Performing differentiation of both sides of \eqref{eq:invmxi_zave} gives
	\begin{equation}
		\frac{\partial \left< |\nabla_\mwc \xi|  \right>_z}{\partial \beta} = \left< U \right>_z \left< |\nabla_\mwc \xi|  \right>_z - \left< U |\nabla_\mwc \xi|  \right>_z \ .
		\label{eq:dinvmxi_dbeta}
	\end{equation}
	Plugging \eqref{eq:dinvmxi_dbeta} into \eqref{eq:dlambxi_dbeta_beginning} yields
	\begin{equation}
		\frac{\partial \ln \left< \lambda_\xi \right>_z}{\partial \beta} = \frac{\kt}{2} + \left< U \right>_z - \frac{\left< U |\nabla_\mwc \xi|  \right>_z}{\left< |\nabla_\mwc \xi|  \right>_z} \ .
		\label{eq:dlambxi_dbeta}
	\end{equation}
	Finally we have from \eqref{eq:P_alpha}
	\begin{align}
		\frac{\partial \ln \mathcal{P}(\mathrm{R})}{\partial \beta} & = \frac{\partial}{\partial \beta} \left( \ln Z_\mathrm{R} - \ln Z \right)
		= \left< U \right> - \left< U \right>_\mathrm{R} \ ,
		\label{eq:dlnP_dbeta}
	\end{align}
	where we deduced from \eqref{eq:Z_molec} 
	\begin{equation}
		\frac{\partial \ln Z_\alpha}{\partial  \beta} = 
		\frac{1}{Z_\alpha} \frac{\partial Z_\alpha}{\partial \beta}
		= - \frac{\int_{\Omega_\alpha} \rmd \bx\ e^{-\beta U(\bx)}\ U(\bx)}{\int_{\Omega_\alpha} \rmd \bx\ e^{-\beta U(\bx)}} 
		= - \left< U \right>_\alpha \ .
		\label{eq:dlnZ_dbeta}
	\end{equation}
	From the definition of the partition function of the unconstrained system follows that
	\begin{equation}
		\frac{\partial \ln Q}{\partial \beta} = - \left< U \right> - 3N\kt / 2 \ .
		\label{eq:dlnQ_dbeta}
	\end{equation}

	\section{Computation of Constrained Ensemble Averages}
	\label{App:Fixman}
	
	As the free-energy, internal-energy, and entropy profiles are defined, they involve ensemble averages over the phase space $\{\bq', \bp_q'\}$ of the constrained system. 
	However, it is very inconvenient in practice to perform the ensemble average under the constraints $q_1=z$ and $p_1=0$. 
	There is an alternate pathway to the constrained ensemble average within the framework of the (Cartesian) phase space $\{\bx, \bp\}$ of the unconstrained system, as we now demonstrate. 
	We start with the expression for the constrained average of a quantity $O(\bx)$ that depends only on the coordinates,
	\begin{equation}
		\left< O \right>_{q_1=z, p_1=0}
		= \frac{1}{Q(z)\ h^{3N-1}} \int \rmd \bq' \int \rmd \bp_q'\ e^{-\beta \mathcal{H}(z)}\ O(z, \bq')
		\label{eq:appB_eq1}
	\end{equation}
	where $Q(z)$  is the constrained partition function (\eqref{eq:Q_z}) and  $\mathcal{H}(z)$ is the constrained Hamiltonian (\eqref{eq:Hamiltonian_z}). 
	Using the same trick that we employed in the transition from \eqref{eq:Q_z} to \eqref{eq:Q_z_expanded}, we obtain from \eqref{eq:appB_eq1}
	\begin{align}
		\left< O \right>_{q_1=z, p_1=0}
		& = \frac{\sqrt{\frac{h^2}{2\pi \kt}}}{Q(z)\ h^{3N}} \nonumber \\
		& \qquad \times \int \rmd \bq \int \rmd \bp_q\ e^{-\beta \mathcal{H}}\ O(\bq)\ |\nabla_\mwc q_1|\ \delta[q_1 - z] \ .
	\end{align}
	Transformation from curvilinear to Cartesian coordinates and performance of the integrations on the momenta yields 
	\begin{align}
		\left< O \right>_{q_1=z, p_1=0} & = 
		\frac{\sqrt{\frac{h^2}{2\pi \kt}} }{Q\ \rho(z)\ \left< \lambda_\xi \right>_z} 
		\prod_i^{3N}  \sqrt{\frac{2\pi m_i \kt}{h^2}} \nonumber \\
		& \qquad \times  \int \rmd \bx\ e^{-\beta U(\bx)}\ O(\bx)\ |\nabla_\mwc \xi|\ \delta[\xi(\bx) - z] \ ,
		\label{eq:appB_eq3}
	\end{align}
	where \eqref{eq:Q_z_short} is used to replace  $Q(z)$. 
	Using the relation $Q = Z/\Lambda$ along with eqs.~(\ref{eq:deBroglie}) and (\ref{eq:Z_complete}), we recast \eqref{eq:appB_eq3} as 
	\begin{align}
		\left< O \right>_{q_1=z, p_1=0} & = 
		\frac{\sqrt{\frac{h^2}{2\pi \kt}}}{\rho(z)\ \left< \lambda_\xi \right>_z} \nonumber \\
		& \qquad \times \frac{\int \rmd \bx\ e^{-\beta U(\bx)}\ O(\bx)\ |\nabla_\mwc \xi|\ \delta[\xi(\bx) - z]}{Z} \nonumber \\
		& = \frac{\sqrt{\frac{h^2}{2\pi \kt}}}{\rho(z)\ \left< \lambda_\xi \right>_z}
		\left< O(\bx)\ |\nabla_\mwc \xi|\ \delta[\xi(\bx) - z]  \right> \ .
	\end{align}
	From eqs.~(\ref{eq:lambda_xi}) and (\ref{eq:m_xi}) we have $\left< \lambda_\xi \right>_z = \sqrt{{h^2} / {2\pi \kt}} \left< |\nabla_\mwc \xi| \right>_z$, which finally gives
	\begin{equation}
		\left< O \right>_{q_1=z, p_1=0} = \frac{\left< O\ |\nabla_\mwc \xi| \right>_z}{\left< |\nabla_\mwc \xi| \right>_z} \ .
		\label{eq:Fixman}
	\end{equation}
	Since $|\nabla_\mwc \xi| = m_\xi^{-1/2} $ one can interpret \eqref{eq:Fixman} as mass-weighted average of the quantity $O$.
	
	\eqref{eq:Fixman} rationalizes the form of the weighted average of the potential energy in the activation internal energy and the internal-energy profile
	\begin{equation}
		\frac{\left< U\ |\nabla_\mwc \xi|  \right>_{z}}{\left< |\nabla_\mwc \xi|  \right>_{z}} = \left< U \right>_{q_1=z, p_1=0} \ .
	\end{equation}
	The relation of the averages over the constrained and unconstrained ensembles (\eqref{eq:Fixman}) was previously derived by Fixman\cite{Fixman1974} and also in the context of the blue moon ensemble\cite{Carter1989}.

	\clearpage
	\onecolumngrid
	\begin{center}
		\LARGE
		Supporting Material for: Entropy and Energy Profiles of Chemical Reactions
	\end{center}

	\setcounter{section}{0}
	\setcounter{figure}{0}
	\renewcommand{\thesection}{S\arabic{section}}
	\renewcommand{\theequation}{S\arabic{equation}}
	\renewcommand\figurename{Figure S\hspace*{-4px}}
	\renewcommand\tablename{Table S\hspace*{-4px}}
	\section{Alternative Derivation of Entropy and Internal Energy Profiles}
	
	The internal energy can be written
	\begin{equation}
		E 
		= \left< \mathcal{H} \right> = \left< K + U \right> = \left< K\right> + \left< U \right>
		\label{eq:E}
	\end{equation}
	The entropy can be expressed as\cite{Hill:1960aa} 
	\begin{equation}
		S = \frac{-\kb}{h^{3N}} \int \rmd \bx\ \int \rmd \bp\ \frac{e^{-\beta \mathcal{H}}}{Q} \ln \frac{e^{-\beta \mathcal{H}}}{Q}
		\label{eq:S}
	\end{equation}
	For the unconstrained system \eqref{eq:E} gives for the inner energy
	\begin{align}
		E & = \frac{\int \rmd \bx\ \int \rmd \bp\ e^{-\beta(K(\bp) + U(\bx))}\ (K(\bp) + U(\bx))}{\int \rmd \bx\ \int \rmd \bp\ e^{-\beta(K(\bp) + U(\bx))}} \nonumber \\
		& = \frac{\int \rmd \bx\  e^{-\beta U(\bx)}\ U(\bx)}{\int \rmd \bx\  e^{-\beta U(\bx)}} + \frac{\int \rmd \bp\ e^{-\beta K(\bp)}\ K(\bp) }{\int \rmd \bp\ e^{-\beta K(\bp)}}  \nonumber \\
		& = \left<U \right> + \sum_{i=1}^{3N} \frac{\sqrt{1/2\pi m_i \kt}}{2m_i}  \int_{-\infty}^\infty \rmd p_i\ e^{-\beta \frac{p_i^2}{2m_i}}\ p_i^2 \nonumber \\
		& = \left<U \right> + \sum_{i=1}^{3N}  \frac{\sqrt{1/2\pi m_i \kt}}{2m_i}  \frac{\sqrt{\pi}}{2} \left(2 m_i \kt \right)^{3/2} \nonumber \\
		& =   \left<U \right> + \sum_{i=1}^{3N} \sqrt{\frac{1}{2\pi m_i \kt}} \frac{\sqrt{\pi}}{4m_i} \left(2 m_i \kt \right)^{3/2} \nonumber \\
		& = \left<U \right> + \frac{3N}{2}\kt
		\label{eq:E_solved}
	\end{align}
	and for the entropy \eqref{eq:S} gives
	\begin{align}
		S 
		& = \frac{-\kb}{h^{3N}Q} \int \rmd \bx\ \int \rmd \bp\ e^{-\beta \mathcal{H}} \left[-\beta \mathcal{H} - \ln Q \right] \nonumber \\
		& = \frac{-\kb}{h^{3N}Q} \int \rmd \bx\ \int \rmd \bp\ e^{-\beta \mathcal{H}}\ \beta \left[F - \mathcal{H} \right] \nonumber \\
		& = \frac{1}{T} \frac{\int \rmd \bx\ \int \rmd \bp\ e^{-\beta \mathcal{H}}\ \left[\mathcal{H} - F  \right]}{h^{3N}Q} \nonumber \\
		& = \frac{1}{T} \left[\left< \mathcal{H} \right> -  F   \right] = \frac{1}{T} \left[\left< U \right> + \frac{3N}{2}\kt - F  \right] 
	\end{align}
	where we use the relation $F=-\kt \ln Q$.
	
	For the constrained system this approach gives
	\begin{align}
		E(z) & = \left< \mathcal{H}(z) \right>_{q_1=z, p_1=0} \nonumber \\
		& = \frac{\int \rmd \bq'\ \int \rmd \bp'\ e^{-\beta \mathcal{H}(z)}\ \mathcal{H}(z)}{\int \rmd \bq'\ \int \rmd \bp'\ e^{-\beta \mathcal{H}(z)}} \nonumber \\
		& = \frac{\int \rmd \bq\ \int \rmd \bp\ e^{-\beta \mathcal{H}}\ \mathcal{H}(z)\ |\nabla_\mwc \xi|\ \delta[q_1 - z] }{\int \rmd \bq\ \int \rmd \bp\ e^{-\beta \mathcal{H}}\ |\nabla_\mwc \xi|\ \delta[q_1 - z] } 
	\end{align}
	We then add and subtract the missing term in the kinetic energy in the numerator yielding:
	\begin{align}
		E(z) & = \frac{\int \rmd \bq\ \int \rmd \bp\ e^{-\beta \mathcal{H}}\ \mathcal{H}\ |\nabla_\mwc \xi|\ \delta[q_1 - z] }{\int \rmd \bq\ \int \rmd \bp\ e^{-\beta \mathcal{H}}\ |\nabla_\mwc \xi|\ \delta[q_1 - z] } - \frac{\int \rmd \bq\ \int \rmd \bp\ e^{-\beta \mathcal{H}}\ \frac{p_1^2}{2m_\xi}\ |\nabla_\mwc \xi|\ \delta[q_1 - z] }{\int \rmd \bq\ \int \rmd \bp\ e^{-\beta \mathcal{H}}\ |\nabla_\mwc \xi|\ \delta[q_1 - z] } 
		\nonumber \\
		& = \frac{\int \rmd \bx\ e^{-\beta U}\ U\ |\nabla_\mwc \xi|\ \delta[\xi(\bx) - z]}{\int \rmd \bx\ e^{-\beta U}\ |\nabla_\mwc \xi|\ \delta[\xi(\bx) - z] } 
		+ \frac{\int \rmd \bp_x\ e^{-\beta K}\ K}{\int \rmd \bp_x\ e^{-\beta K}} 
		- \frac{\sqrt{2\pi (\kt)^3}/2}{\sqrt{2\pi \kt}}\frac{\int \rmd \bq\ \int \rmd \bp'\ e^{-\beta \mathcal{H}}\ \delta[q_1 - z] }{\int \rmd \bq\ \int \rmd \bp'\ e^{-\beta \mathcal{H}}\  \delta[q_1 - z] }
	\end{align}
	The second line was achieved by transformation of the first term to Cartesian coordinates and separation of momenta and coordinates. In the second term an integration over $p_1$ was carried out.
	One can see that the integrals in the last term cancel. 
	The first term in eq.~(S6) is expanded with $1 = \frac{\int \rmd \bq\ e^{-\beta U}\ \delta[q_1 - z]}{\int \rmd \bq\ e^{-\beta U}\ \delta[q_1 - z]}$ such that numerator and denominator can be rewritten as $z$-conditioned ensemble averages.
	\begin{align}
		E(z)
		& =  \frac{\left<U\ |\nabla_\mwc \xi| \right>_z }{\left<|\nabla_\mwc \xi| \right>_z} + \frac{3N\kt}{2} - \frac{\kt}{2} \nonumber \\
		& = \frac{\left<U\ |\nabla_\mwc \xi| \right>_z }{\left<|\nabla_\mwc \xi| \right>_z} + \frac{3N-1}{2}\kt
	\end{align}
	The constrained entropy profile can alternatively be derived as:
	\begin{align}
		S(z) & = \frac{-\kb}{h^{3N-1}} \int \rmd \bq'\ \int \rmd \bp'\ \frac{e^{-\beta \mathcal{H}(z)}}{Q(z)} \ln \frac{e^{-\beta \mathcal{H}(z)}}{Q(z)} \nonumber \\
		& = \frac{-\kb}{h^{3N-1}Q(z)} \int \rmd \bq'\ \int \rmd \bp'\ e^{-\beta \mathcal{H}(z)} \left[-\beta \mathcal{H}(z) - \ln Q(z) \right] \nonumber \\
		& = \frac{-\kb}{h^{3N-1}Q(z)} \int \rmd \bq'\ \int \rmd \bp'\ e^{-\beta \mathcal{H}(z)}\ \beta \left[F(z) - \mathcal{H}(z) \right] \nonumber \\
		& = \frac{-1}{T} \left< F(z) \right>_z  + \frac{1}{T} \left< \mathcal{H}(z) \right>_{q_1=z, p_1=0} \nonumber \\
		& = \frac{1}{T} \left( - F(z) + \frac{\left<U\ |\nabla_\mwc \xi| \right>_z }{\left<|\nabla_\mwc \xi| \right>_z} + \frac{3N-1}{2}\kt \right)
	\end{align}	
	Note that these results are identical to eqs.~(33) and~(34) of the article.
	
	
	\section{Relation Between FEPs Whose CVs Have Parallel Gradients}
	\label{app:Invertible}
	
	Here we provide derivations of the relation between FEPs, whose CVs possess parallel gradients in configuration space.
	A key operation required during the following development is transformation from massweighted coordinates $\{\mwc\}$ to curvilinear coordinates $\{\mathbf{q} \}$ and vice versa. 
	Performance of this operation requires an expression for the Jacobian associated with the transformation. 
	According to Wong and York\cite{Wong2012}, the magnitude of the determinant of the Jacobian matrix for the transformation from mass-weighted coordinates to curvilinear coordinates $\{ \xi, \mathbf{q}'\}$ is given by
	\begin{equation}
		|\mathbf{J}_\xi| = \prod_{i=1}^{3N} |\nabla_\mwc q_i|^{-1}
		\label{eq:Jacbian_xi}
	\end{equation}
	For the transformation from mass-weighted to curvilinear coordinates $\{ \phi, \mathbf{q}'\}$, we have by analogy with \eqref{eq:Jacbian_xi}
	\begin{equation}
		|\mathbf{J}_\phi| = |\nabla_\mwc \phi|^{-1} \ \prod_{i=2}^{3N} |\nabla_\mwc q_i|^{-1}
		\label{eq:Jacobian_phi}
	\end{equation}
	We first examine the relation between the (constrained) free-energy profiles (FEP) corresponding to the two CVs. 
	Rewriting eq.~(28) in terms of mass-weighted coordinates, we have
	\begin{equation}
		Q^\xi(z) = C \ \int \rmd \mwc\ e^{-\beta U(\mwc)} |\nabla_\mwc \xi|\ \delta[\xi(\mwc) - z] 
		\label{eq:Q_xi_parallelCVs}
	\end{equation}
	where we define the constant by $C \equiv Q \sqrt{h^2/2\pi\kt}/Z$.
	The superscript $\xi$ on $Q^\xi(z)$ signifies the CV to which the FEP corresponds. 
	By an analogous derivation we obtain the FEP for the \enquote{parallel} CV
	\begin{equation}
		Q^\phi(z') = C \ \int \rmd \mwc\ e^{-\beta U(\mwc)} |\nabla_\mwc \phi|\ \delta[\phi(\mwc) - z'] 
		\label{eq:Q_phi_parallelCVs}
	\end{equation}
	where C is the same constant as that defined in \eqref{eq:Q_xi_parallelCVs}. 
	Note that we place a prime on the fixed value of the CV $\phi$, in order to distinguish it from the fixed value of the CV $\xi$ in \eqref{eq:Q_xi_parallelCVs}. 
	The parameters are related by
	\begin{equation}
		z = f^{-1}(z') \quad \mathrm{or} \quad f(z) = z'
		\label{eq:z_inverse_zprime}
	\end{equation}
	Transforming from mass-weighted to curvilinear coordinates $\{ \phi, \mathbf{q}'\}$ in \eqref{eq:Q_phi_parallelCVs}, we obtain
	\begin{align}
		Q^\phi(z') & = C \int \rmd \phi\ \int \rmd \mathbf{q}'\ |\mathbf{J}_\phi|\ e^{-\beta U(\mathbf{q})}\ |\nabla_\mwc \phi|\ \delta[\phi - z'] \nonumber \\
		& = C \int \rmd \mathbf{q}'\ \prod_{i=2}^{3N} |\nabla_\mwc q_i|^{-1}\ e^{-\beta U(z', \mathbf{q}')}\ 
		\label{eq:Q_phi_intermediate}
	\end{align}
	where the second line was obtained by substitution with \eqref{eq:Jacobian_phi} and integration over $\phi$.
	Eq.~(\ref{eq:Q_phi_intermediate}) can be recast as
	\begin{align}
		Q^\phi(z') & = C \int \rmd q_1 \int \rmd \mathbf{q}'\  \nonumber \\
		& \qquad \frac{|\nabla_\mwc q_1|}{|\nabla_\mwc q_1|}\ \prod_{i=2}^{3N} |\nabla_\mwc q_i|^{-1}\ e^{-\beta U(z', \mathbf{q}')}\ \delta[q_1 - f^{-1}(z')] \nonumber \\
		& = C \int \rmd q_1 \int \rmd \mathbf{q}'\ \nonumber \\
		& \qquad |\mathbf{J}_\xi|\ e^{-\beta U(z', \mathbf{q}')}\ |\nabla_\mwc q_1|\ \delta[q_1 - f^{-1}(z')]
		\label{eq:Q_phi_intermediate2}
	\end{align}            
	where the second line of \eqref{eq:Q_phi_intermediate2} uses eqs.~(\ref{eq:Jacbian_xi}) and~(\ref{eq:z_inverse_zprime}). 
	Finally, transformation back to massweighted coordinates gives
	\begin{equation}
		Q^\phi(z') = C \int \rmd \mwc\ e^{-\beta U(\mwc)}\ |\nabla_\mwc \xi|\ \delta[\xi - f^{-1}(z')] = Q^\xi(f^{-1}(z'))
		\label{eq:Q_phi_final}
	\end{equation}
	Now comparing eqs.~(\ref{eq:Q_phi_final}) and (\ref{eq:Q_xi_parallelCVs}), we deduce the following relation
	\begin{align}
		F^\phi(z') & = F^\xi(z) = F^\xi(f^{-1}(z'))
	\end{align}

	We turn next to a derivation of the relation between \enquote{standard} PMFs that correspond to CVs having parallel gradients. 
	That is, we desire to determine the analogue of \eqref{eq:Q_phi_final} for the PMF.
	The probability density for the CV $\xi$ is
	\begin{align}
		\rho^\xi(z) & = \frac{1}{Z} \int \mathrm{d}\bx\ e^{-\beta U(\bx)}\ \delta[\xi(\bx) - z] \nonumber \\
		& = \frac{\prod_i m_i^{-1/2}}{Z} \int \mathrm{d}\mwc\ e^{-\beta U(\mwc)}\ \delta[\xi(\mwc) - z] 
	\end{align}
	The probability density for the CV $\phi$ is
	\begin{align}
		\rho^\phi(z') & =  \frac{\prod_i m_i^{-1/2}}{Z} \int \mathrm{d}\mwc\ e^{-\beta U(\mwc)}\ \delta[\phi(\mwc) - z'] \nonumber \\
		& = C \int \mathrm{d}\phi\ \mathrm{d}\mathbf{q}'\ |J_\phi|\ e^{-\beta U(\mwc)}\ \delta[\phi - z']\nonumber \\
		& = C \int \mathrm{d}\phi\ \mathrm{d}\mathbf{q}'\ |\nabla_\mwc f(q_1)|^{-1}\ \prod_{i=2}^{3N}|\nabla_\mwc q_i|^{-1}\ e^{-\beta U(f(q_1), \mathbf{q}')}\ \delta[f(q_1) - z'] \nonumber \\
		& = C \int \mathrm{d}\mathbf{q}'\ |f'(f^{-1}(z'))|^{-1}|\nabla_\mwc q_1|^{-1}\ \prod_{i=2}^{3N}|\nabla_\mwc q_i|^{-1}\ e^{-\beta U(z', \mathbf{q}')} \nonumber \\
		& = C \int \mathrm{d}\mathbf{q}\ |f'(f^{-1}(z'))|^{-1}\ |J_\xi|\ e^{-\beta U(\mathbf{q})}\ \delta[q_1 - f^{-1}(z')] \nonumber \\
		& = C \int \mathrm{d}\mwc\ |f'(f^{-1}(z'))|^{-1}\  e^{-\beta U(\mwc)}\ \delta[\xi - f^{-1}(z')] \nonumber \\
		& = |f'(f^{-1}(z'))|^{-1}\ C \int \mathrm{d}\mwc\   e^{-\beta U(\mwc)}\ \delta[\xi - f^{-1}(z')] \nonumber \\
		\rho^\phi(z') & =  |f'(f^{-1}(z'))|^{-1}\ \rho^\xi(f^{-1}(z'))  \ ,
	\end{align}    	
	where we have implicitly defined $C = \frac{\prod_i m_i^{-1/2}}{Z}$
	
	To summarize, we have
	\begin{align}
		\rho^\xi(z) & = \rho^\phi(f(z))\ \left|f'(z) \right| \ ,
	\end{align}
	from which it follows by eq.~(43) of the manuscript that
	\begin{align}
		A^\xi(z) & = A^\phi(f(z)) - \kb T \ln |f'(z)|
	\end{align}
	We conclude that it is possible for two PMFs whose CVs have parallel gradients to assume different values on the same iso-surface.

	\section{Pseudo Inverse of Curvilinear Gradient}
	\label{sec:pseudoinverse}
	
	Let $\bx$ denote a point in configuration space. 
	$x_1, x_2, \dots, x_n$ denote the Cartesian coordinates, $\widetilde{x}_1, \widetilde{x}_2, \dots, \widetilde{x}_n$ mass-weighted coordinates, and $q_1, q_2, \dots, q_n$ a set of curvilinear coordinates.
	The basis vectors in Cartesian coordinates can be expressed as
	\begin{equation}
		e_i = \frac{\partial \bx}{\partial x_i} \ ,
	\end{equation}
	The unit basis vectors in mass-weighted coordinates are identical to the Cartesian basis vectors (the non-unit mass-weighted basis vectors do not have unit length but are parallel to the Cartesian basis vectors).
	\begin{equation}
		e_i = \frac{1}{\left|\frac{\partial \bx}{\partial \widetilde{x}_i} \right|} \frac{\partial \bx}{\partial \widetilde{x}_i} = \frac{\partial \widetilde{x}_i}{\partial x_i} \frac{\partial \bx}{\partial \widetilde{x}_i} = \frac{\partial \mwc}{\partial \widetilde{x}_i}
	\end{equation}
	%
	%
	The basis vectors from one basis can be transformed into  another one as
	\begin{equation}
		e_i = \frac{\partial \mwc}{\partial \widetilde{x}_i} = \sum_j \frac{\partial q_j}{\partial x_i} \frac{\partial \mwc}{\partial q_j}
		= \sum_j \frac{\partial q_j}{\partial \widetilde{x}_i} \frac{\partial \mwc}{\partial q_j}
		\label{eq:basis_expansion}
	\end{equation}
	We assume that at least the first curvilinear coordinate $q_1$ is orthogonal with respect to mass-weighted coordinates to all other curvilinear coordinates, as in the derivation of $F(z)$ in the article (see \eqref{eq:orthogonality_crit}).
	
	The mass-weighted gradient of the first curvilinear coordinate is
	\begin{equation}
		\nabla_\mwc q_1 = \sum_i \frac{\partial q_1}{\partial \widetilde{x}_i} e_i \ .
		\label{eq:mass_weighted_gradient}
	\end{equation}
	Inserting \eqref{eq:basis_expansion} in \eqref{eq:mass_weighted_gradient} yields
	\begin{align}
		\nabla_\mwc q_1 & = \sum_i \frac{\partial q_1}{\partial \widetilde{x}_i} e_i \\
		& = \sum_i \frac{\partial q_1}{\partial \widetilde{x}_i} \sum_j \frac{\partial q_j}{\partial \widetilde{x}_i} \frac{\partial \mwc}{\partial q_j} \\
		& = \sum_j  \frac{\partial \mwc}{\partial q_j} \sum_i \frac{\partial q_1}{\partial \widetilde{x}_i} \frac{\partial q_j}{\partial \widetilde{x}_i} \\
		& = \sum_j \frac{\partial \mwc}{\partial q_j} \left( \nabla_\mwc q_1 \cdot \nabla_\mwc q_j \right)
	\end{align}
	Using the orthogonality from eq.~(24) from the article gives
	\begin{equation}
		\nabla_\mwc q_1 = \frac{\partial \mwc}{\partial q_1} |\nabla_\mwc q_1|^2
	\end{equation}
	The orthogonality assumption from the article is
	\begin{equation}
		(\nabla_\mwc q_1)^\mathrm{T}\ \nabla_\mwc q_i 
		= (\nabla_\mwc \xi)^\mathrm{T}\ \nabla_\mwc q_i 
		= \delta_{1i} |\nabla_\mwc \xi|^2 \ ,
		\label{eq:orthogonality_crit}
	\end{equation}
	Rearranging the terms finally gives
	\begin{equation}
		\frac{\partial \mwc}{\partial q_1} = \frac{\nabla_\mwc q_1}{|\nabla_\mwc q_1|^2}
	\end{equation}
	This means that under the assumption of \eqref{eq:orthogonality_crit} the partial derivative with respect to $q_1$ of any scalar function $f$ can be expressed as
	\begin{equation}
		\frac{\partial f}{\partial q_1} = \frac{\partial f}{\partial \mwc}\cdot \frac{\partial \mwc}{\partial q_1}=  \nabla_\mwc f \cdot \frac{\nabla_\mwc q_1}{|\nabla_\mwc q_1|^2}
		\label{eq:pseudoinverse}
	\end{equation}

	\section{Derivation of the Expression for the Mean Constrained Force}
	
	From eqs.~(30) and (31) of the article, we have
	\begin{align}
		- \frac{\partial F(z)}{\partial z} & = \kt \frac{\partial }{\partial z} \ln Q(z) = \frac{\kt}{Q \rho(z) \left< \lambda_\xi \right>_z} \frac{\partial }{\partial z} \left[ Q \rho(z) \left< \lambda_\xi \right>_z \right] \nonumber \\
		& = \frac{\kt}{\rho(z) \left< \lambda_\xi \right>_z} \frac{\partial }{\partial z} \left[ \rho(z) \left< \lambda_\xi \right>_z \right] \nonumber \\
		& = \frac{\kt}{\rho(z) \left< \lambda_\xi \right>_z} \frac{\partial }{\partial z} \left[ \frac{\int \rmd \bx\ e^{- \beta U}\ \delta[\xi(\bx) -z]}{Z} \frac{\int \rmd \bx\ e^{- \beta U}\ \delta[\xi(\bx) -z]\ \lambda_\xi(\bx)}{\int \rmd \bx\ e^{- \beta U}\ \delta[\xi(\bx) -z]} \right] \nonumber \\
		& = \frac{\kt}{\rho(z) \left< \lambda_\xi \right>_z} \frac{\partial }{\partial z} \left[ \frac{\int \rmd \mwc\ e^{- \beta U}\ \delta[\xi(\mwc) -z]}{\sqrt{\prod_i m_i}\ Z} \frac{\int \rmd \mwc\ e^{- \beta U}\ \delta[\xi(\mwc) -z]\ \lambda_\xi(\mwc)}{\int \rmd \mwc\ e^{- \beta U}\ \delta[\xi(\mwc) -z]} \right] \nonumber \\
		& = \frac{\kt}{\rho(z) \left< \lambda_\xi \right>_z} \frac{\partial }{\partial z} \left[ \frac{\int \rmd \mwc\ e^{- \beta U}\ \delta[\xi(\mwc) -z]\ \lambda_\xi(\mwc)}{\sqrt{\prod_i m_i}\ Z} \right] \nonumber \\
		& = \frac{\kt}{\sqrt{\prod_i m_i}\ Z \rho(z) \left< |\nabla_{\mwc} \xi| \right>_z}  \left[ \frac{\partial }{\partial z} \int \rmd \mwc\ e^{- \beta U}\ \delta[\xi(\mwc) -z]\ |\nabla_{\mwc} \xi| \right] \nonumber \\
		& = \frac{\kt}{\sqrt{\prod_i m_i}\ Z \rho(z) \left< |\nabla_{\mwc} q_1| \right>_z}  \left[ \frac{\partial }{\partial z} \int \rmd \bq\ |\mathbf{J}|\ e^{- \beta U}\ \delta[q_1 -z]\ |\nabla_{\mwc} q_1| \right] \nonumber \\
		& = \frac{\kt}{\sqrt{\prod_i m_i}\ Z \rho(z) \left< |\nabla_{\mwc} q_1| \right>_z}  \left[ \frac{\partial }{\partial z} \int \rmd \bq'\ |\mathbf{J}(z, \bq')|\ e^{- \beta U(z, \bq')}\ |\nabla_{\mwc} q_1|_{q_1=z} \right] \nonumber \\
		& = \frac{\kt}{\sqrt{\prod_i m_i}\ Z \rho(z) \left< |\nabla_{\mwc} q_1| \right>_z}  \left[  \int \rmd \bq'\ \frac{\partial |\mathbf{J}|}{\partial q_1} \ e^{- \beta U( \bq)}\ |\nabla_{\mwc} q_1|
		+ |\mathbf{J}| \frac{\partial e^{- \beta U( \bq)}}{\partial q_1} |\nabla_{\mwc} q_1|
		+|\mathbf{J}|\ e^{- \beta U(\bq)}\ \frac{\partial |\nabla_{\mwc} q_1|}{\partial q_1} \right] \nonumber \\
		& = \frac{\kt}{Z \rho(z) \left< |\nabla_{\mwc} q_1| \right>_z}   \int \rmd \bx\ \delta[\xi(\bx) - z]\ e^{- \beta U( \bq)}\ \left[ \frac{1}{|\mathbf{J}|}\frac{\partial |\mathbf{J}|}{\partial q_1} \ |\nabla_{\mwc} q_1|
		- \beta \frac{\partial  U( \bq)}{\partial q_1} |\nabla_{\mwc} q_1|
		+ \frac{\partial |\nabla_{\mwc} q_1|}{\partial q_1} \right] \nonumber \\
		&=\frac{1}{\left< |\nabla_\mwc q_1| \right>_z} \left<  |\nabla_{\mwc} q_1|\ \left( - \frac{\partial U(\bq)}{\partial q_1} \right)
		+ \kt \frac{|\nabla_{\mwc} q_1|}{|\mathbf{J}|} \frac{\partial |\mathbf{J}|}{\partial q_1}
		+ \kt \frac{\partial |\nabla_{\mwc} q_1|}{\partial q_1} \right>_z
		\label{eq:dF_dz_long}
	\end{align}
	For the derivative of a scalar function with respect to the first curvilinear coordinate $q_1$ we use \eqref{eq:pseudoinverse} derived in the previous Section.
	
	The gradient of the length of a gradient can be expressed as the product of gradient and Hessian
	\begin{align}
		\nabla_\mwc |\nabla_\mwc q_1| & = \frac{1}{|\nabla_\mwc q_1|} \left( 
		\begin{array}{c}
			\sum_{i} \left(\frac{\partial q_1}{\partial \widetilde{x}} \right) \left(\frac{\partial^2 q_1}{\partial \widetilde{x}_i\partial \widetilde{x}_1}\right) \\
			\vdots
			\\
			\sum_{i} \left(\frac{\partial q}{\partial \widetilde{x}_i}\right) \left(\frac{\partial^2 q_1}{\partial \widetilde{x}_i\partial \widetilde{x}_{3N}}\right)
		\end{array}
		\right) \nonumber \\
		& = \frac{1}{|\nabla_\mwc q_1|} \left(\nabla_\mwc q_1 \right)^T \widetilde{\mathbf{H}}(q_1) \ ,
	\end{align}
	where $\widetilde{\mathbf{H}}(q_1)$ is the Hessian matrix of $q_1$ with respect to mass-weighted coordinates.
	In analogy to Ref.~\cite{denOtter2000}, where Cartesian derivatives are used, we take the derivative of the Jacobian matrix to be
	\begin{equation}
		\frac{1}{|\mathbf{J}|}\frac{\partial |\mathbf{J}|}{\partial q_1} = 
		\nabla_\mwc \cdot  \frac{\nabla_\mwc q_1}{|\nabla_\mwc q_1|^2} = \frac{1}{|\nabla_\mwc q_1|^4} \left[ \Delta_\mwc q_1 |\nabla_\mwc q_1|^2 - 2 \left(\nabla_\mwc q_1\right)^T \widetilde{\mathbf{H}}(q_1) \nabla_\mwc q_1 \right]
	\end{equation}    
	The derivative of the length of the mass-weighted gradient of $q_1$ is
	\begin{align}
		\frac{\partial |\nabla_{\mwc} q_1|}{\partial q_1} 
		& = \nabla_{\mwc} |\nabla_{\mwc} q_1| \cdot \frac{\nabla_{\mwc} q_1}{|\nabla_{\mwc} q_1|^2} \nonumber \\
		& = \left[ \frac{1}{|\nabla_{\mwc} q_1|} \left(\nabla_{\mwc} q_1\right)^T \widetilde{\mathbf{H}}(q_1) \right] \cdot \frac{\nabla_{\mwc} q_1}{|\nabla_{\mwc} q_1|^2} \nonumber \\
		& = \frac{\left(\nabla_{\mwc} q_1\right)^T \widetilde{\mathbf{H}}(q_1) \nabla_{\mwc} q_1}{|\nabla_{\mwc} q_1|^3} 
	\end{align}
	In order to add $\frac{\partial |\nabla_{\mwc} q_1|}{\partial q_1}$ to the Jacobian derivative $|\nabla_{\mwc} q_1|\left(\frac{1}{|J|} \frac{\partial |J|}{\partial q_1}\right)$ we expand the former by multiplying it with $\frac{|\nabla_{\mwc} q_1|}{|\nabla_{\mwc} q_1|}$.
	\begin{align}
		\frac{|\nabla_{\mwc} q_1|}{|\nabla_{\mwc} q_1|}\frac{\partial |\nabla_{\mwc} q_1|}{\partial q_1} & = |\nabla_{\mwc} q_1| \left( \frac{1}{|\nabla_{\mwc} q_1|} \frac{\partial |\nabla_{\mwc} q_1|}{\partial q_1} \right) \nonumber \\
		& = |\nabla_{\mwc} q_1| \left(\frac{\left( \nabla_{\mwc} q_1\right)^T \widetilde{\mathbf{H}}(q_1) \left( \nabla_{\mwc} q_1 \right)}{|\nabla_{\mwc} q_1|^4} \right) 
	\end{align}
	Combining the derivative of the Jacobian and the length of the gradient yields
	\begin{align}
		\frac{1}{|J|} \frac{\partial |J|}{\partial q_1} +  \frac{1}{|\nabla_{\mwc} q_1|} \frac{\partial |\nabla_{\mwc} q_1|}{\partial q_1} 
		& = \frac{\left[ \Delta_\mwc q_1 |\nabla_\mwc q_1|^2 - 2 \left(\nabla_\mwc q_1\right)^T \widetilde{\mathbf{H}}(q_1) \nabla_\mwc q_1 \right] + \left( \nabla_\mwc q_1\right)^T \widetilde{\mathbf{H}}(q_1) \ \nabla_\mwc q_1}{|\nabla_\mwc q_1|^4}  \nonumber \\
		& = \frac{ \Delta_\mwc q_1 |\nabla_\mwc q_1|^2 -  \left(\nabla_\mwc q_1\right)^T \widetilde{\mathbf{H}}(q_1) \nabla_\mwc q_1  }{|\nabla_\mwc q_1|^4}  \nonumber \\
		& = \frac{ \nabla_\mwc \cdot \nabla_\mwc q_1 }{|\nabla_\mwc q_1|^2}   - \frac{ (\nabla_\mwc |\nabla_\mwc q_1|) \cdot \nabla_\mwc q_1 }{|\nabla_\mwc q_1|^3} \nonumber \\
		& = \frac{1}{|\nabla_\mwc q_1|} \left[ \frac{|\nabla_\mwc q_1| \nabla_\mwc \cdot \nabla_\mwc q_1 }{|\nabla_\mwc q_1|^2}   - \frac{ (\nabla_\mwc |\nabla_\mwc q_1|) \cdot \nabla_\mwc q_1 }{|\nabla_\mwc q_1|^2} \right] \nonumber \\
		& = \frac{1}{|\nabla_\mwc q_1|} \left[ \nabla_\mwc \cdot \frac{\nabla_\mwc q_1}{|\nabla_\mwc q_1|} \right] \nonumber \\
	\end{align}
	Inserting these results into \eqref{eq:dF_dz_long} gives
	\begin{align}
		- \frac{\partial F(z)}{\partial z} 
		& = \frac{1}{\left< |\nabla_\mwc q_1| \right>_z} \left<  |\nabla_{\mwc} q_1|\ \left( - \frac{\partial U(\bq)}{\partial q_1} \right)
		+ \kt \frac{|\nabla_{\mwc} q_1|}{|J|} \frac{\partial |J|}{\partial q_1}
		+ \kt \frac{\partial |\nabla_{\mwc} q_1|}{\partial q_1} \right>_z \nonumber \\
		& = \frac{1}{\left<| \nabla_\mwc q_1 |\right>_z} \left<  |\nabla_{\mwc} q_1|\ \left( - \nabla_\mwc U \cdot \frac{\nabla_\mwc q_1  }{|\nabla_\mwc q_1|^2} + \frac{\kt}{|\nabla_\mwc q_1|} \left[ \nabla_\mwc \cdot \frac{\nabla_\mwc q_1}{|\nabla_\mwc q_1|} \right] \right)
		\right>_z \\
		& = \frac{1}{\left< |\nabla_\mwc q_1| \right>_z} \left< - \nabla_\mwc U \cdot \frac{\nabla_{\mwc} q_1  }{|\nabla_{\mwc} q_1|} + {\kt} \left[ \nabla_\mwc \cdot \frac{\nabla_{\mwc} q_1}{|\nabla_{\mwc} q_1|} \right]
		\right>_z
	\end{align}

	After having derived an explicit form of the mean constrained force, we aim to obtain the analogous expression for the mean force to have both derivations in the same notation.
	Expressions for the mean force have already been derived previously.\cite{denOtter2000, Henin2004}
	\begin{align}
		- \frac{\partial A(z)}{\partial z} & = \kt \frac{\partial }{\partial z} \ln \rho(z) = \frac{\kt}{\rho(z)} \frac{\partial }{\partial z} \rho(z) \nonumber \\
		& = \frac{\kt}{\rho(z)} \frac{\partial }{\partial z} \frac{\int \rmd \bx\ e^{- \beta U}\ \delta[\xi(\bx) -z]}{Z} \nonumber \\
		& = \frac{\kt}{Z\ \rho(z)} \frac{\partial }{\partial z} \int \rmd q_1\ \rmd\bq'\ |\mathbf{J}|\ e^{- \beta U}\ \delta[q_1 -z] \nonumber \\
		& = \frac{\kt}{Z\ \rho(z)} \frac{\partial }{\partial q_1} \int \rmd\bq'\ |\mathbf{J}|\ e^{- \beta U}\ \nonumber \\
		& = \frac{\kt}{Z\ \rho(z)} \int \rmd\bq'\ |\mathbf{J}|\ e^{- \beta U}\ \left[- \beta \frac{\partial U}{\partial q_1} + \frac{1}{|\mathbf{J}|}\frac{\partial |\mathbf{J}|}{\partial q_1} \right] \nonumber \\
		& = \frac{\kt}{Z\ \rho(z)} \int \rmd\bq\ |\mathbf{J}|\ e^{- \beta U}\ \left[- \beta \frac{\partial U}{\partial q_1} + \frac{1}{|\mathbf{J}|}\frac{\partial |\mathbf{J}|}{\partial q_1} \right]\ \delta[q_1 -z] \nonumber \\
		& = \frac{\kt}{Z\ \rho(z)} \int \rmd\bx\ e^{- \beta U}\ \left[- \beta \frac{\partial U}{\partial q_1} + \frac{1}{|\mathbf{J}|}\frac{\partial |\mathbf{J}|}{\partial q_1} \right]\ \delta[q_1 -z] \nonumber \\
		& = \left<- \frac{\partial U}{\partial q_1} + \kt \frac{1}{|\mathbf{J}|}\frac{\partial |\mathbf{J}|}{\partial q_1} \right> 
	\end{align}
	The derivation above was carried out starting from Cartesian coordinates and using the assumption that $\nabla q_1 \cdot \nabla q_j = \delta_{1j} |\nabla q_1|^2$.
	However, it can also be done from mass-weighted coordinates and using $\nabla_\mwc q_1 \cdot \nabla_\mwc q_j = \delta_{1j} |\nabla_\mwc q_1|^2$, which would lead to an analogous expression.
	Therefore, both of the following expressions are arguably equal.
	\begin{align}
		- \frac{\partial A(z)}{\partial z} & = \left< - \nabla U \cdot \frac{\nabla q_1  }{|\nabla q_1|^2} + {\kt} \left[ \nabla \cdot \frac{\nabla q_1}{|\nabla q_1|^2} \right]
		\right>_z \\
		& = \left< - \nabla_\mwc U \cdot \frac{\nabla_{\mwc} q_1  }{|\nabla_{\mwc} q_1|^2} + {\kt} \left[ \nabla_\mwc \cdot \frac{\nabla_{\mwc} q_1}{|\nabla_{\mwc} q_1|^2} \right]
		\right>_z 
	\end{align}

	\section{Argument that $A(z)$ is a PMF}
	\label{App:ProofPMF}
	
	The PMF is sometimes \cite{denOtter2000} defined by adding the total free energy of the system to the expression for $A(z)$ given by eq.~(43) of the article. 
	\begin{align}
		A(z) & = F - \kt \ln \rho(z) = - \kt \ln [ Q\ \rho(z)] \nonumber \\
		& = - \kt \ln \frac{Z}{\Lambda} \frac{\int \rmd \bx\ e^{-\beta U(\bx)}\ \delta[\xi(\bx)-z]}{Z} \nonumber \\
		& = - \kt \ln \frac{1}{h^{3N}} \int \rmd \bx \int \rmd \bp\ e^{-\beta \mathcal{H}(\bx, \bp)}\ \delta[\xi(\bx)-z] \nonumber \\
		& = - \kt \ln Q^\star(z)
	\end{align}
	The pseudo-partition function in the last line is starred to distinguish it from the constrained partition function $Q(z)$ in eq.~(27) of the article.
	The gradient of this extended PMF with respect of the value of the CV gives:
	\begin{align}
		-\frac{\rmd A(z)}{\rmd z} & = \frac{\kt}{Q^\star(z)} \frac{\partial Q^\star(z)}{\partial z} \nonumber \\
		& = \frac{\kt}{Q^\star(z)} \frac{\partial }{\partial z} \frac{1}{h^{3N}} \int \rmd \bp_q \int \bq\ e^{-\beta \mathcal{H}(q_1, \bq', \bp_q)}\ \delta[q_1 - z]\nonumber \\
		& = \frac{\kt}{Q^\star(z)} \frac{1}{h^{3N}} \int \rmd \bp_q \int \bq'\ \frac{\partial }{\partial z} e^{-\beta \mathcal{H}(z, \bq', \bp_q)}\nonumber \\
		& = \frac{-1}{Q^\star(z)} \frac{1}{h^{3N}} \int \rmd \bp_q \int \bq'\ e^{-\beta \mathcal{H}(z, \bq', \bp_q)}\ \frac{\partial  \mathcal{H}}{\partial z} \nonumber \\
		& = \left< -\frac{\partial  \mathcal{H}}{\partial z} \right>_{q_1=z} \label{eq:mean_force}\\
		& = \left< -\frac{\partial  U}{\partial z} - \frac{\partial  K}{\partial z} \right>_{q_1=z}
		\label{eq:mean_force_split}
	\end{align}
	$\left< \ \right>_{q_1=z}$ signifies the ensemble average over all momenta and all coordinates but $q_1=z$.
	\eqref{eq:mean_force} shows that $A(z)$ is the potential of the mean instantaneous force.
	Comparison of eq.~(45) of the article and \eqref{eq:mean_force_split} explains that the first term in eq.~(45) stems from the PES and the second term, which is said to be of geometric character, is the derivative of the kinetic energy. 
	The difference between \eqref{eq:mean_force_split} and eq.~(36) is the extra term in the kinetic energy.
	
	\section{Contact Pair}
	\label{ape:contact_pair}
	Here we treat the contact pair in detail.
	
	Since the solvent is homogeneous, properties of the pair do not depend on the position $\mathbf{R}$ of the center of mass. 
	Thus, we take $\mathbf{R}$ to be fixed and, consequently, $\mathbf{P}=0$. 
	Lastly, we assume that the accessible relative configuration space is contained in a ball of radius $r_\mathrm{max}$. 
	We take the CV to be
	\begin{equation}
		\xi(x, y, z) = \sqrt{x^2 + y^2 + z^2} = r
	\end{equation}
	where $x$, $y$, and $z$ are the Cartesian components of the distance between the two atoms. 
	The symmetry of the system dictates use of spherical coordinates, in terms of which the Hamiltonian (for the relative motion) assumes the form
	\begin{equation}
		\mathcal{H} = \frac{p_r^2}{2\mu} + \frac{p_\theta^2}{2\mu r^2} + \frac{p_\phi^2}{2\mu r^2 \sin^2 \theta} + U(r)
	\end{equation}
	where $\mu$ is the reduced mass of the contact pair and $U(r)$ is the effective PES mediated by the solvent.
	Imposing the constraint that $\xi = r = z$ is fixed, so that $p_r = 0$, we obtain the constrained Hamiltonian
	\begin{equation}
		\mathcal{H}(z) = \frac{p_\theta^2}{2\mu z^2} + \frac{p_\phi^2}{2\mu z^2 \sin^2 \theta} + U(z)
		\label{eq:cp_Hz}
	\end{equation}
	where $z$ now stands for the fixed value of the CV. We can now calculate the constrained partition function directly from its formal expression
	\begin{align}
		Q(z) & = \frac{1}{h^2} \int_0^\pi \rmd \theta \int_0^{2\pi} \rmd \phi \int_{-\infty}^\infty \rmd p_\theta \int_{-\infty}^\infty \rmd p_\phi  e^{-\beta \mathcal{H}(z)} \nonumber \\
		& = \frac{4\pi z^2 e^{-\beta U(z)}}{\lambda^2}
	\end{align}
	where the thermal de Broglie wavelength for this case is
	\begin{equation}
		\lambda = \sqrt{\frac{h^2}{2\pi\mu \kt}}
	\end{equation}


	\section{Profiles for the Cyclization of the Hexenyl Radical}
	
	Ref.~\cite{DietschreitDiestlerBombarelli2022} describes how the unbiased Boltzmann weights for each simulation frame were obtained.
	The mass $m_\xi$ for the CV $\xi = d(\mathrm{C1-C5})$ is the reduced mass of the two atoms (roughly 6~a.m.u.), which remains constant throughout the simulation. 
	Therefore, $E(z)$ is computed as a simple $z$-conditioned ensemble average of the PES along the range of CV values, $E(z) = \left<U\right>_z$. 
	The IEP is more noisy than the FEP and is smoothed by convolution with a Gaussian kernel (\texttt{scipy.ndimage.gaussian\_filter1d}) with a standard deviation of 1.5 bin widths, 0.075~\AA.
	The FEP differs from the PMF by a constant. 
	The zero point is set  at the equilibrium position of the C-C bond.
	The entropy profile is $S(z) = \frac{1}{T} (E(z) - F(z))$, where $T$ is 300~K.
	
	\section{Profiles for LGPS}
	
	\begin{figure*}[bt]
		\centering
		\includegraphics[width=0.49\linewidth]{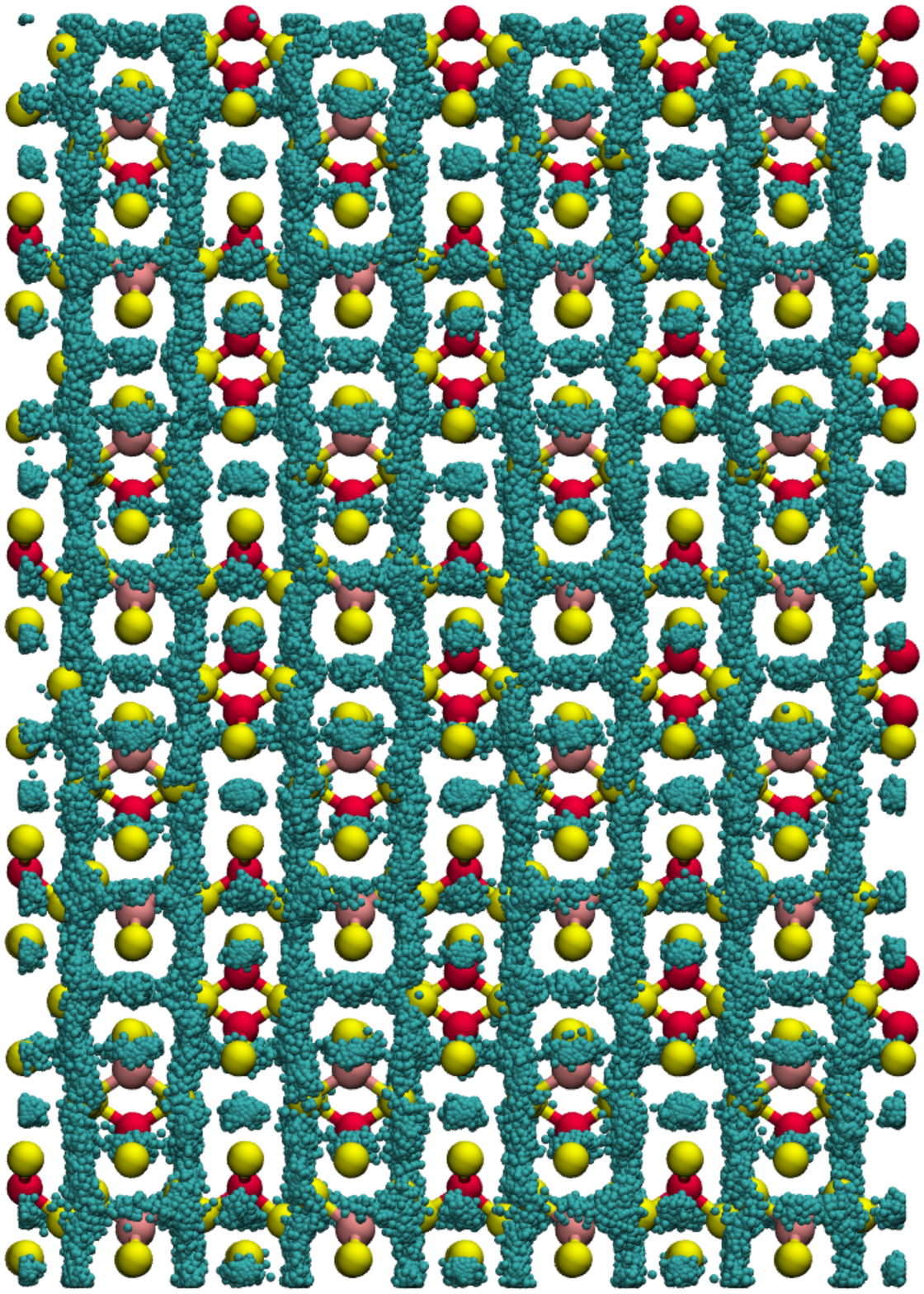}
		\includegraphics[width=0.49\linewidth]{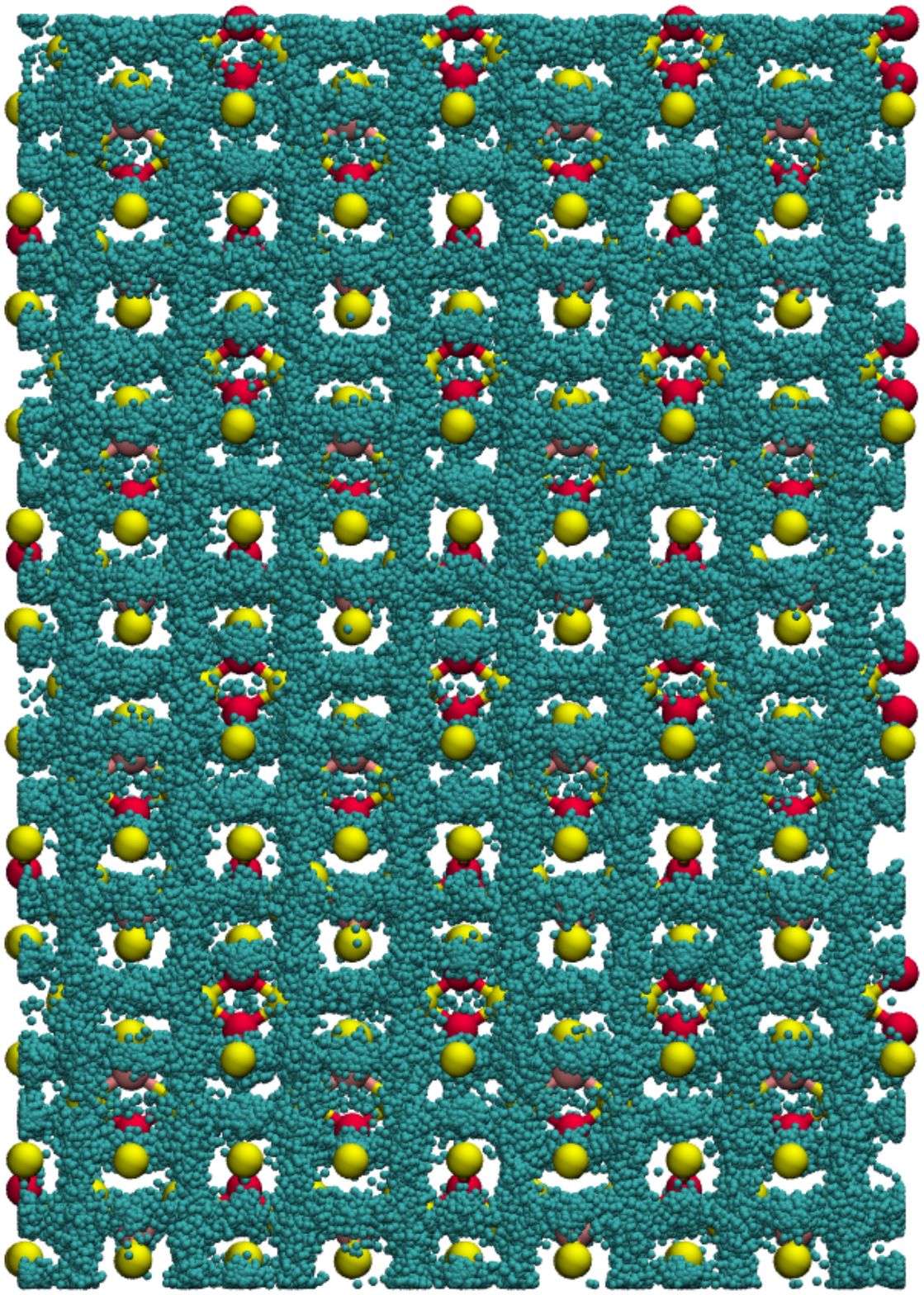}
		\caption{Orthographic visualization of the $yz$-plane of the LGPS simulation cell at 400~K (left) and 800~K (right).
			The Li ions are shown in cyan as overlays from many MD frames.
			The positions of the phosphorous (red), sulfur (yellow), and germanium (salmon) atoms are taken from the first MD frame.}
		\label{fig:LGPS_cell}
	\end{figure*}
	
	As stated in the article, the simulation data were made available to us by the authors of Ref.~\cite{Gavin2023}.
	The simulations are 1~ns long and were carried out with a time step of 1~fs in the NVT ensemble. 
	Atomic configurations were written to disk every 5 time steps.
	We discarded the first 10~ps of each simulation as equilibration period. 
	Each simulation box contains 1280 lithium atoms, as the box contains 64 unit cells, with 4 unit cells along each edge of the box.
	
	Figure~S1 shows an overlay of simulation frames at 400~K and 800~K.
	The positions of the sulfur, germanium, and phosphor atoms are shown only for the initial frame and lithium positions as an overlay for equidistant time steps.
	The overlay of Li$^+$ positions in the left part of Fig.~S1 shows fully connected channels along the vertical axis (conductive channels along z-axis) and disconnected Li$^+$ pockets in the horizontal direction.
	At 800~K (right part of Fig.~S1), conduction  occurs along all three axes as can be seen from the much increased spread of lithium.

	In order to compute profiles for a single unit cell, the x-, y-, and z-coordinates of every lithium atom were taken with respect to the origin to the unit cell in which the Li atom occurred.
	Averaged histograms were constructed yielding the probability density of finding a lithium atom along each axis of a unit cell.
	Since the CV is a Cartesian coordinate of lithium, $m_\xi = 6.941$~a.m.u.
	Thus, for example, the average FEP was obtained by 
	\begin{equation}
		\hat{F}_\alpha (z) = - \kt \ln \sum_{i=1}^{N_\mathrm{ion}} \frac{\rho_{\alpha i}(z)}{N_\mathrm{ion}}
	\end{equation}
	where $\alpha$ denotes the Cartesian component and $i$ labels the ions.
	
	The internal-energy profiles (IEP, $E_{x,y,z}(z)$) are constructed as histograms of the potential energy, by analogy with the construction of the probability densities. 
	For each MD frame the potential energy for each of the 1280 Li atoms was binned. 
	The potential energy of such a large system varies over a range of hundreds of kJ/mol. 
	However, inclusion of all Li atoms in constructing the histogram, and not just a single one, yields meaningful IEPs.  
	Note that we do not include the contribution from the mean kinetic energy, since it is constant at        $3(N-1)\kt/2$. 
	The IEP was smoothed by convolution with a Gaussian kernel with a standard deviation of one bin width, 0.1~\AA.
	
	Error bars were obtained by splitting the MD trajectories into 5 parts and performing the analysis for each part separately. 
	The error along the profiles is then computed as the standard deviation of the five profiles obtained form the subsampling for each property.
	Like the FEP for the cyclization reaction, the FEP for migration of Li ions shows no visible error.

	\section{Computational Details of Chabazite Simulations}
	
	The simulations were carried out with a PaiNN \cite{Schutt2021} model within the ASE  framework.
	The NN potential was trained in \cite{Millan2023} and made available to us.
	Simulations were carried for three different configurations of the system. 
	Each system consisted of a 2x2x4 chabazite cell, containing one aluminum atom on a silicon site and a [Cu(NH$_3$)$_2$]$^+$ complex.
	The three systems differ in the placement of the aluminum atom as the movement of the copper complex is constrained to a particular range along the CV.
	If the simulations were unconstrained, all three systems would be identical due to the placement of the aluminum at analogous silicon sites and periodicty of the box.
	The CV was the same for all simulations, as described in the article. 
	The simulations were carried out under periodic boundary conditions with our own ASE implementation of the WTM-eABF \cite{lesage2017smoothed, fu2018zooming, fu2019taming} algorithm.
	The system was confined by a harmonic potential to a CV range of -10.4 to +10.4~\AA, which effectively traps the complex along the axis that contains 4 unit cells.
	Motion orthogonal to the CV was unconfined. 
	
	For each of the three different aluminum positions, 40 simulations with a length of 2.002~ns were carried out. 
	The time step was 0.5~fs. 
	The first 2~ps of each simulation were discarded as equilibration phase. 
	Thereby a total of 80~ns of production simulations are amassed for each system.
	The temperature was maintained with the Langevin thermostat using a friction coefficient of 1~ps$^{-1}$.
	
	The unbiased weights of each configuration were obtained with MBAR\cite{Hulm2022}.
	The FEP and IEP were computed according to eqs.~(30) and~(34) of the article, respectively. 
	The mass-weighted gradient of the complicated CV was computed via automatic differentiation in pyTorch. 
	The bin width for the profiles is 0.2~\AA.
	The IEP was smoothed by convolution with a Gaussian kernel with a standard deviation of 0.3~\AA.
	The error for each profile was computed by splitting the total data from the 80~ns randomly into 10 parts.
	The analysis was carried out for each fraction individually and the error was computed as the standard deviation of the ten profiles. 
	
	\begin{table}
		\centering
		\caption{Reaction and activation free- and internal energies and entropies for migration of the Cu complex in chabazite. 
			Energy in units of kJ/mol and entropy in units of J/mol~K. 
			Numbers after $\pm$ are standard deviations, computed as detailed above.}
		\label{tab:chab_values}
		{
			\begin{tabular}{r| r r | r r | r r}
				\multicolumn{1}{c}{} & \multicolumn{2}{c}{Fig.~3a} & \multicolumn{2}{c}{Fig.~3b} & \multicolumn{2}{c}{Fig.~3c} \\
				$X$ & $\Delta X$ & $\Delta X^\ddagger$ & $\Delta X$ & $\Delta X^\ddagger$ & $\Delta X$ & $\Delta X^\ddagger$ \\
				$F$ & $0.6 \pm 0.0$ & $32.2 \pm 0.1$  & $-31.0 \pm 0.0$ & $24.1 \pm 0.1$  & $-10.9 \pm 0.0$ & $20.2 \pm 0.0$ \\
				$E$ & $1.8 \pm 0.2$ & $18.9 \pm 1.9$  & $-41.5 \pm 0.3$ & $8.2 \pm 1.5$   & $-7.1 \pm 0.4$  & $14.5 \pm 1.8$ \\
				$S$ & $3.0 \pm 0.4$ & $-31.4 \pm 4.6$ & $-24.7 \pm 0.6$ & $-37.5 \pm 3.5$ & $9.1 \pm 0.9$   & $-13.4 \pm 4.2$
			\end{tabular}
		}
	\end{table}

	\begin{figure}
		\centering
		\includegraphics[width=0.49\linewidth]{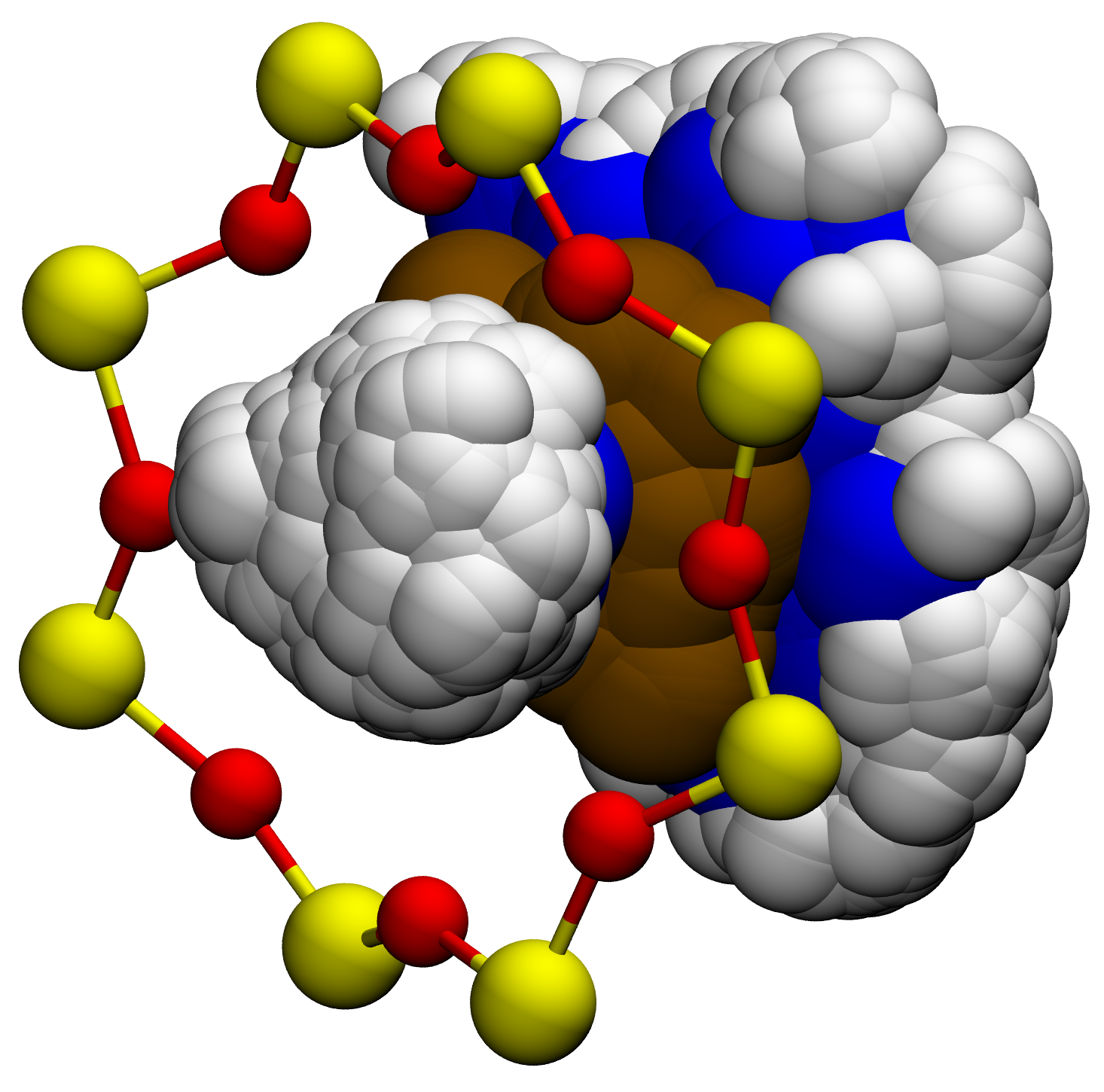}
		\includegraphics[width=0.49\linewidth]{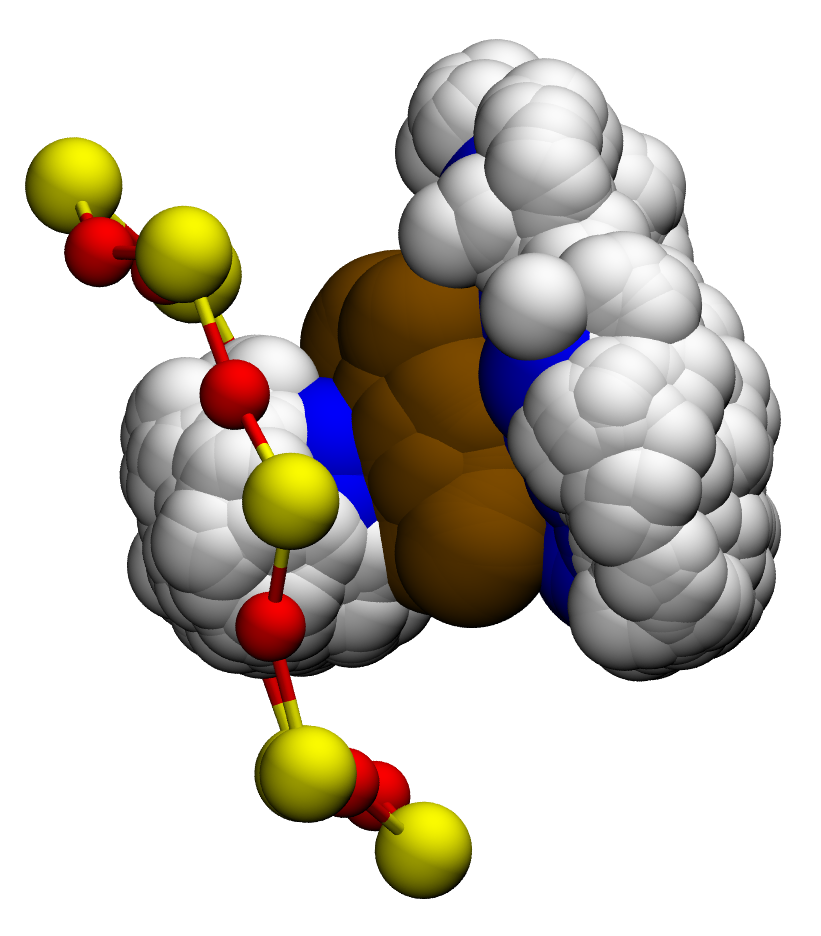}
		\includegraphics[width=0.49\linewidth]{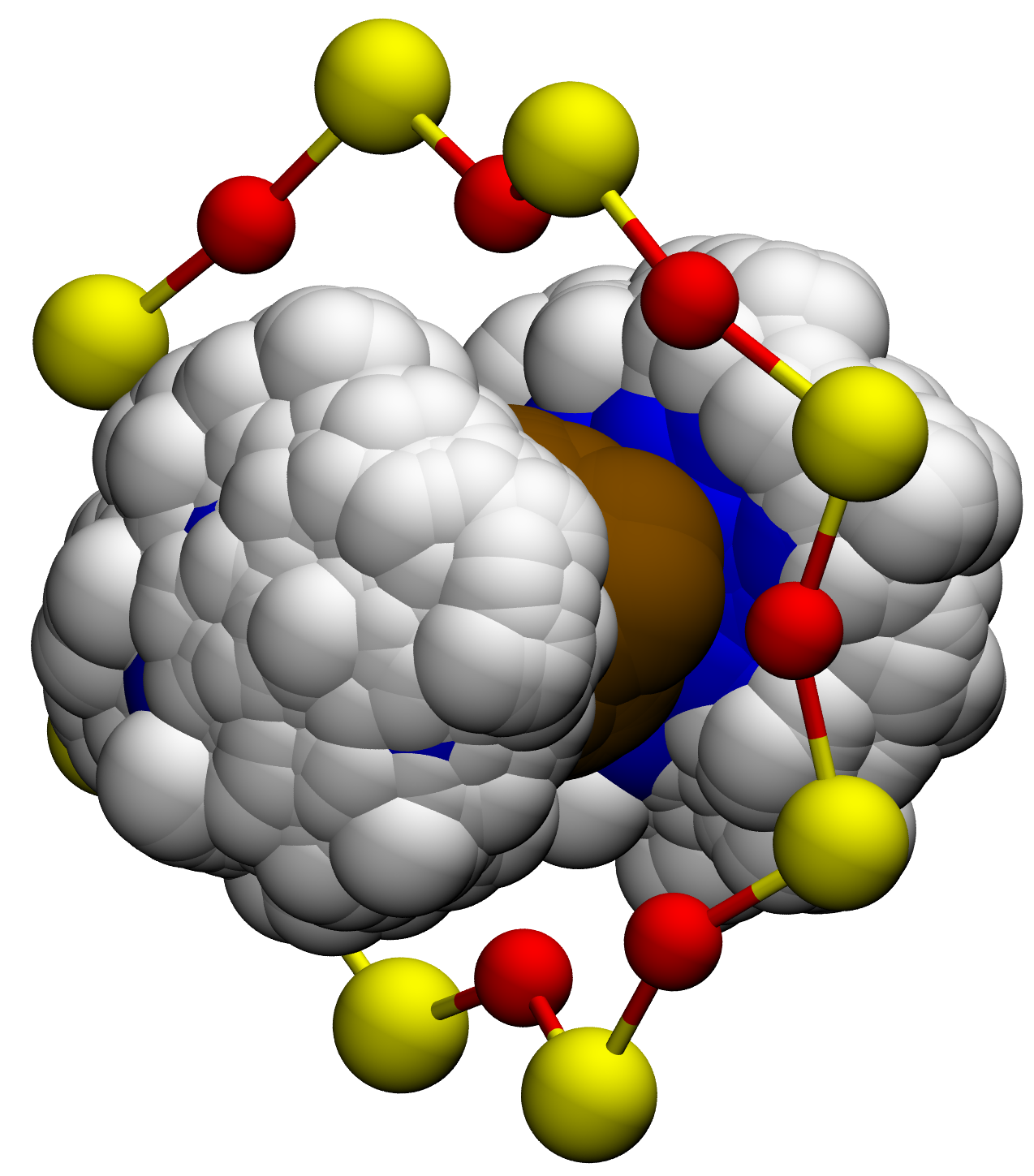}
		\includegraphics[width=0.49\linewidth]{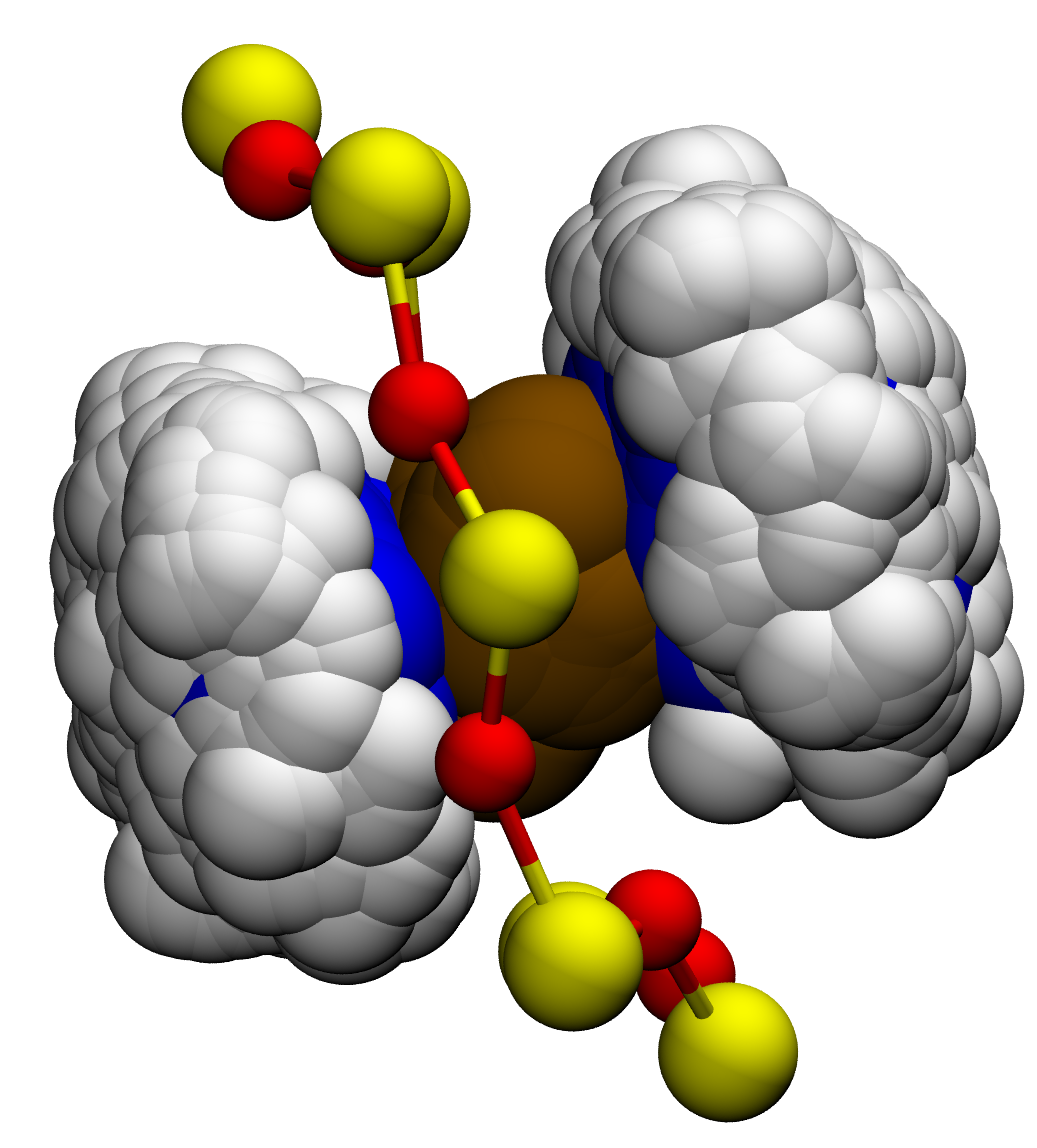}
		\caption{Visualizations of Cu-complex configurations from two different angles taken from the simulations with the aluminum far removed. 
			In the top row the configurations have CV values close to -2, corresponding to the minimum in the inner energy left of the barrier in Fig.~3a in the article. 
			The second row shows configurations of the same system with CV values close to the transition point value (-0.3). 
			For better visibility only one configuration is shown for the 8-ring.
			The decrease in entropy when the ammonia hydrogen atoms form hydrogen bonds is visible.}
		\label{fig:my_label}
	\end{figure}

\end{document}